\documentstyle[11pt,aaspp4,flushrt]{article}

\newcommand{\degr}{\hbox{$^\circ$}}

\lefthead{M.~Kramer et al.}
\righthead{The characteristics of millisecond pulsar emission I.}

\begin{document}

\title{The characteristics of millisecond pulsar emission: \\
I.~Spectra, pulse shapes and the beaming fraction}

\author{Michael Kramer\altaffilmark{1}, 
Kiriaki M.~Xilouris\altaffilmark{2},
Duncan R.~Lorimer\altaffilmark{1},
Oleg Doroshenko\altaffilmark{1},
Axel Jessner\altaffilmark{1},
Richard Wielebinski\altaffilmark{1},
Alexander Wolszczan\altaffilmark{3},
Fernando Camilo \altaffilmark{4}}

\altaffiltext{1}{Max-Planck-Institut f\"ur Radioastronomie, Auf dem H\"ugel 69,
53121  Bonn, Germany} 
\altaffiltext{2}{National Astronomy and Ionosphere Center, Arecibo Observatory, P.O. Box 995, Arecibo, PR 00613, USA}
\altaffiltext{3}{Department of Astronomy and Astrophysics, 
Penn State University, University Park, PA 16802, USA} 
\altaffiltext{4}{Nuffield Radio Astronomy Laboratories,
Jodrell Bank, Macclesfield, Cheshire SK11 9DL, England; Marie Curie Fellow} 

\begin{abstract}

The extreme physical conditions in millisecond pulsar
magnetospheres as well as their different evolutionary history
compared to ``normal pulsars'' raise the question as to 
whether these objects also differ in their radio emission properties.
We have monitored a large sample of millisecond
pulsars for a period of three years using the 100-m Effelsberg
radio telescope in order to compare
the radio emission properties of these two pulsar populations.
Our sample comprises a homogeneous data set of very high quality.

With some notable exceptions,
our findings suggest that the two groups 
of objects share many common properties.
A comparison of the spectral indices between samples 
of normal and millisecond
pulsars demonstrates that millisecond pulsar spectra
are not significantly different from those of normal pulsars. This is
contrary to what has previously been thought. There is evidence,
however, that millisecond pulsars are slightly less luminous 
and less efficient radio emitters compared to normal pulsars.
We confirm recent suggestions that a 
diversity exists among the luminosities of millisecond pulsars with the
isolated millisecond pulsars
being less luminous than the binary millisecond
pulsars, implying
an influence of the different evolutionary history on the emission properties.
There are indications that old millisecond pulsars exhibit somewhat
flatter spectra than the presumably younger ones.

Contrary to common belief, we present evidence that the millisecond pulsar
profiles are only marginally more complex than those found among the normal pulsar
population. Moreover, the development of the profiles with frequency is
rather slow, suggesting  very compact magnetospheres. The profile development
seems to anti-correlate with the companion mass and the
spin period, again suggesting that the amount
of mass transfer in a binary system might directly influence the 
emission properties. The angular radius of
radio beams of millisecond pulsars does not follow the scaling predicted
from a canonical pulsar model which is applicable for normal pulsars.
Instead they are systematically smaller, supporting the concept
of a critical rotational period below which such a scaling 
ceases to exist.
The smaller inferred luminosity and narrower emission beams will need to be
considered in 
future calculations of the birth-rate of the Galactic population.

\end{abstract}
\keywords{Pulsars: millisecond pulsars -- normal pulsars -- sub-millisecond pulsars
-- emission mechanism -- birth-rates}

\section{Introduction}

At the time of the discovery of the first millisecond pulsar B1937+21
by Backer et
al.~(1982) around 350 pulsars were known. While most of these had periods
around one second, the period of PSR B1937+21 was a mere 1.6 ms, 
immediately raising the question in what respect such
objects differ from long period pulsars. Like the original binary pulsar,
B1913+16 with a period of 59 ms, it 
was proposed that millisecond pulsars originate in
binary systems in which a slowly rotating neutron star
gets spun-up by mass transfer from the binary companion (Alpar et
al.~1982). This idea is supported by the fact that 80\%
of all millisecond pulsars found in the Galactic plane are members of binary systems.
These binary systems can be divided into three groups, depending
on the mass of the companion star (Camilo 1996), {\it viz:} low-mass
(presumably Helium) white dwarfs, 
high-mass (presumably Carbon-Oxygen) white dwarfs and other 
neutron stars. The progenitors
are therefore thought to be either low-mass or high-mass 
X-ray binary systems.
It seems fairly well established that millisecond pulsars
are recycled pulsars originating from interacting binary systems. 
As the exact definition of ``millisecond pulsars'' in terms of period
is somewhat arbitrary, it seems reasonable to treat all recycled pulsars as
one class, referring to them as MSPs in the following.

Since the special evolutionary history distinguishes MSPs
from long period pulsars, it is consequently interesting to compare
the observed properties of these two classes of objects.
The most obvious difference between the two classes 
is, of course, the observed range of pulse periods. 
The average period of slowly rotating pulsars 
(hereafter referred to as `normal' pulsars) is about 
0.7 s. Rapidly rotating pulsars exhibit pulse
periods between 1.5 ms and about 60 ms. 
The period derivatives of MSPs
are a few orders of magnitude smaller than those of normal
pulsars, i.e.,~of the order of $10^{-19}$ or smaller compared to
typically $10^{-15}$ for normal pulsars. 
The derived characteristic ages, $\tau= P /2\dot{P}$,
are  of the order of $10^9$ yr for MSPs and thus much larger 
than $10^6$ yr typically found for normal pulsars, 
indicating that MSPs represent
an old population. Similarly, calculating the surface magnetic
field for MSPs, i.e.,~$B=3.2\cdot10^{19} \sqrt{P \dot{P}}$ Gauss 
if $P$ is measured in seconds (Manchester \&
Taylor 1977), we find typically values of the order of only
$10^8$ to $10^9$ Gauss. An outstanding mystery is how the magnetic
field decreases from $10^{12}$ Gauss for normal pulsars by
three to four orders of magnitude during or after the recycling process
(e.g.,~Bhattacharya \& van den Heuvel 1991; Phinney \& Kulkarni 1994).

For normal pulsars, the magnetic field seems to maintain a dipolar
form even at distances of a few stellar radii above the neutron star
surface (Phillips 1992; Xilouris et al.~1996; Kramer
et al.~1997). For MSPs it has been suggested that flux expulsion as a result
of a pre-recycling spin-down (e.g., Srinivasan et al.~1990; Jahan Miri
\& Bhattacharya 1994), or the spin-up process of accreting mass on the neutron star 
surface itself (e.g.,~Romani 1990; Geppert \& Urpin 1996) could not only reduce the 
magnetic field strength, but could also have an impact on the magnetic field 
structure by disturbing the previously present dipolar field.
Moreover, near the surface of MSPs, magnetic multipoles might be 
present (cf.~Ruderman 1991; Arons 1993)
which could increase the actual magnetic field strength 
over the $10^8$ Gauss inferred from the spin down rates
which are dominated by the field structure near the light cylinder (cf.~Krolik 1991).

Despite these differences in the observational properties
of MSPs and normal pulsars, it is often believed that the emission
mechanism responsible for the observed radio emission is the same 
for both types of pulsars (e.g.,~Manchester 1992; Gil \& Krawczyk
1997). 
However, there has been no systematic comparison of the
emission properties of MSPs and normal pulsars. 
In a series of papers, we focus on this question
which also constrains models of the
emission mechanism of normal pulsars. If the same emission theory
applies for normal pulsars and MSPs, it implies that the responsible 
radiation process has to work over three to four orders of magnitude
in spin period and magnetic field, ruling out models which
depend very much on these parameters (see Melrose 1992). In fact, even
if the emission process turns out to be the same for MSPs and normal pulsars,
the special evolutionary history of MSPs can be expected to have a certain
impact on the observed emission properties, so that it is important to separate
such possible effects.

In the past, relatively little work has been done to compare spectral properties 
of MSPs and
normal pulsars. This was primarily due to the small size of the sample
--- only 5 MSPs were known in the Galactic disk prior to 1990.
The first careful study of MSP spectra was
presented by Foster, Fairhead, \& Backer (1991), 
obtaining multi-frequency flux densities for four 
objects. Recently, large-area surveys at Parkes, Arecibo, Jodrell Bank and Green
Bank have been very successful at finding MSPs and, as a
result, over 35 are presently known in the Galactic disk.
Lorimer et al.~(1995b) collected data for 20 MSPs mainly from the
literature and compared them with a sample of normal pulsars, while
Bailes et al.~(1997) compared the luminosities of isolated and binary MSPs. 
Navarro \& Manchester (1996) presented
high-quality data for PSR J0437$-$4715 and demonstrated that MSP profiles
can exhibit an unusually large number of components. As pointed out
by Backer (1995), a systematic study of the
complexity of MSP profiles however has not yet been undertaken. Similarly, due to the limited 
resolution in many of the previously published MSP  profiles, it has been difficult
to investigate the general width of MSP profiles, which is necessary when making
inferences about the intrinsic size of the emission beam. 

In this paper we present data obtained over 
three years during a monitoring project of MSPs at 1400 MHz. 
We used the Effelsberg 100-m radio telescope of the
MPIfR and could thus observe all MSPs with declinations greater 
than $-29\degr$ which were strong enough to obtain data of high
quality. As a result, we present profiles of 27 MSPs, 
of which 24 are located in the Galactic disk, so that our sample
contains about 60\% of all known Galactic disk MSPs. 
Most of the profiles are of high quality both in signal-to-noise ratio
and time resolution, thus allowing much more detailed studies than previously
possible. Therefore, in this and subsequent papers 
(Xilouris et al.~1998a, hereafter paper II; 
Xilouris, Kramer \& von Hoensbroech~1998b, hereafter
paper III; Kramer et al.~1998) we investigate
the largest homogeneous sample of
MSP data studied to date.

We investigate the spectra and observed luminosity distributions
 of MSPs. 
Attempting to infer the structure and size of the emission beam,
we  then  study the profile properties of MSPs. Searching for 
possible differences, a detailed comparison with the properties of normal 
pulsars is made (cf.,~Manchester \& Taylor 1977; Rankin 1983;
Lyne \& Manchester 1988; Taylor, Manchester, \& Lyne 1993;  Malofeev et al.~1994;
Lorimer et al.~1995a; 
Seiradakis et al.~1995, Xilouris et al.~1996, Kijak et al.~1997b). 
This study
is complemented by the subsequent papers mentioned above to 
investigate the polarisation properties of MSPs, the stability of their pulse
profiles and the emission height of their radio emission. 


\section{Instrumental set-up and data reduction}

The data presented here were obtained during a project which
was initiated to obtain pulse time-of-arrival measurements for
a number of MSPs during the upgrade
of the Arecibo telescope\footnote{The Arecibo telescope is
operated by Cornell University for the National Science Foundation.}
 (Wolszczan 1996, Kramer et al.~1996). Since April 1994
we have made regular observations
using the 100-m radio telescope 
in Effelsberg near Bonn. In order to study the polarisation 
characteristics, all four Stokes parameters were recorded.

We used a highly sensitive 1.4 GHz HEMT receiver installed in the prime
focus of the telescope, which is tunable between 1.3 GHz and 1.7 GHz.
The set noise temperature of this system is 25 K at median elevations. 
The antenna gain
at these frequencies is 1.5 K/Jy, independent of elevation.
We obtained left-hand (${\rm LHC} \equiv A$) and right-hand 
(${\rm RHC} \equiv B$) 
circularly polarised  signals across a 
40 MHz bandwidth usually centered at 1.41 GHz, but sometimes
also tuned to 1.71 GHz. The  LHC and RHC signals were mixed down to an
intermediate frequency of 150 MHz and fed into a passive (`adding')
polarimeter providing the complex (undetected) outputs of $A$, $B$,
$A-B$ and $A-iB$ (for a detailed description of this polarimeter and its
calibration see Hoensbroech \& Xilouris 1997b). 
Each output was then split into 60 contiguous channels, each
of 666 kHz bandwidth, by four $60 \times 0.666 = 40$ MHz filter banks.
The individually detected filter outputs were delayed
according to a pre-calculated delay time of the signals caused 
by dispersion due to the interstellar medium. Finally, the filter
outputs were added on-line and these incoherently de-dispersed signals
were transferred to the Effelsberg Pulsar Observing System (EPOS)
backend. 

All four data streams 
could be sampled by the backend with a maximum resolution of
0.2 $\mu$s. However, the pulse period was typically divided into 1024 
phase bins. 
The data were folded according to the computed topocentric
pulse period which was updated every five seconds. Until
the beginning of 1995 the topocentric period was 
calculated using {\sc Tempo} (Taylor \& Weisberg 1989), while later
on we used {\sc Timapr} developed by
Doroshenko \& Kopeikin (1995)
for computational convenience. A
sub-integration of 15 s was finally transferred to disk for off-line
reduction.

In order to monitor the gain stability and polarisation
characteristics of EPOS, we performed regular calibration measurements
using a switchable noise diode. The signal from this noise diode was
injected into the waveguide following the antenna horn 
and was itself
compared to the flux density of known continuum calibrators (Ott 
et al.~1994) during regularly performed pointing observations. 
Switching on the noise diode regularly after observations
of pulsars allowed us to monitor gain differences in LHC and RHC signal
paths and to calibrate the pulse energies by comparing these to 
the noise diode output. Using this reliable method, we obtained
a large number of flux density measurements presented in the following
section.

\section{Millisecond pulsar flux densities}

\label{fluxes}

In Table \ref{fluxtab} we present the results of flux density
measurements for 23 pulsars. These measurements were performed 
repeatedly at 1.41
and 1.71 GHz, and therefore we quote an effective frequency 
$\nu_{\rm eff}$
representing the weighted mean of the centre frequencies used. The error
introduced by this procedure is at most 6\% (for a steep spectral index 
of $-3$) and thus {\it is} much smaller than the uncertainties due to
interstellar scintillation. 
In order to give an impression about the 
amplitude of any remaining uncertainties 
due to scintillation, we additionally quote the median
of the measured flux densities which can be directly compared to the mean
value. 

In the following we compare the spectra and luminosities of MSPs and
normal pulsars. As we will see,
one has to take into account
that the sample of MSPs was solely discovered during 
low frequency surveys, which favoured nearby MSPs
due to severe dispersion and scattering limitations. In contrast,
the sample of normal pulsars includes a large number of
sources found in the previous high
frequency surveys by Johnston et al.~(1992) and Clifton et al.~(1992), 
which in turn favoured flat spectrum, high luminosity sources
(Lorimer et al.~1995b). In order
to construct a more homogeneous set of sources which can then be compared,
we can restrict the studied samples to
those pulsars whose distance is not larger than 1.5 kpc, since we can
reasonably assume that for a comparison of normal pulsars and MSPs 
the population of MSPs has been sufficiently
sampled out to this distance (Lyne et al.~1998). 
Additionally, it will be useful to compare the results
obtained for this set of data with those of an analysis
of MSPs and normal pulsars detected in a single survey. Evidently, 
the recent Parkes 436 MHz southern sky survey 
(Manchester et al.~1996; Lyne et al.~1998)
provides such a homogeneous sample to verify our results.

\subsection{Spectra of MSPs}

\label{indexsect}

Using the flux densities which we have measured at 1.4/1.7 GHz presented
in Table~\ref{fluxtab}, we have used power law fits to calculate spectral 
indices based on our 
flux densities and those published for lower frequencies. Corresponding references
and resulting spectral indices are given in Table~\ref{fluxtab}. We also
included data presented by Foster et al.~(1991), and the recent
observations of four MSPs (J1022+1001, J1713+0747, B1855+09 and 
J2145$-$0750) at 4.85 GHz by Kijak et al.~(1997a). 
If estimated uncertainties
for flux densities found in the literature were not specified, we
assumed a typical error of at least 30\% due to the expected severe
scintillation effects. 
In cases where we believe that scintillation imposes an even larger 
uncertainty, we accounted for
this effect by increasing the estimated error. 
Additionally, for those pulsars which are not included in our sample
but discussed in the literature, we derived 
spectral indices for 11 further MSPs located in the Galactic disk
(second part of Table~\ref{fluxtab}).
Our subsequent analysis does not 
include PSRs J2019+2425 and J2322+2057,
although we present pulse profiles for these sources.
For both these objects the number of flux density
measurements was too small to derive a reliable mean value for 1.4/1.7 GHz. Presently 
only flux densities at 430 MHz are given in the literature 
(Nice, Taylor, \& Fruchter~1993; Nice \& Taylor 1995).

In order to compare the spectra for 
Galactic disk MSPs and normal pulsars, 
we used a set of spectral indices for 346 normal pulsars 
which was derived from data
presented by Malofeev et al.~(1994), Lorimer et al.~(1995b) and also
Taylor et al.~(1995). We find that the average
spectrum of MSPs seems to be steeper than that
of normal pulsars, i.e. we derive a mean spectral index
of $-1.8\pm0.1$  for a set of 32 MSPs located in the Galactic disk 
and a mean index of $-1.60\pm0.04$ for normal
pulsars in the same frequency range.
\footnote{Errors quoted here correspond to the mean standard deviation
of the mean.}
The median values for both samples
are $-1.8$ and $-1.7$, respectively. 
Although the mean spectrum for MSPs seems to be somewhat steeper, 
both distributions appear very similar. 
A Kolmogorov-Smirnov (KS) test yields
a probability of 36\% that they are drawn from the same parent 
distribution.

\begin{figure}
\epsscale{0.5}
\plotone{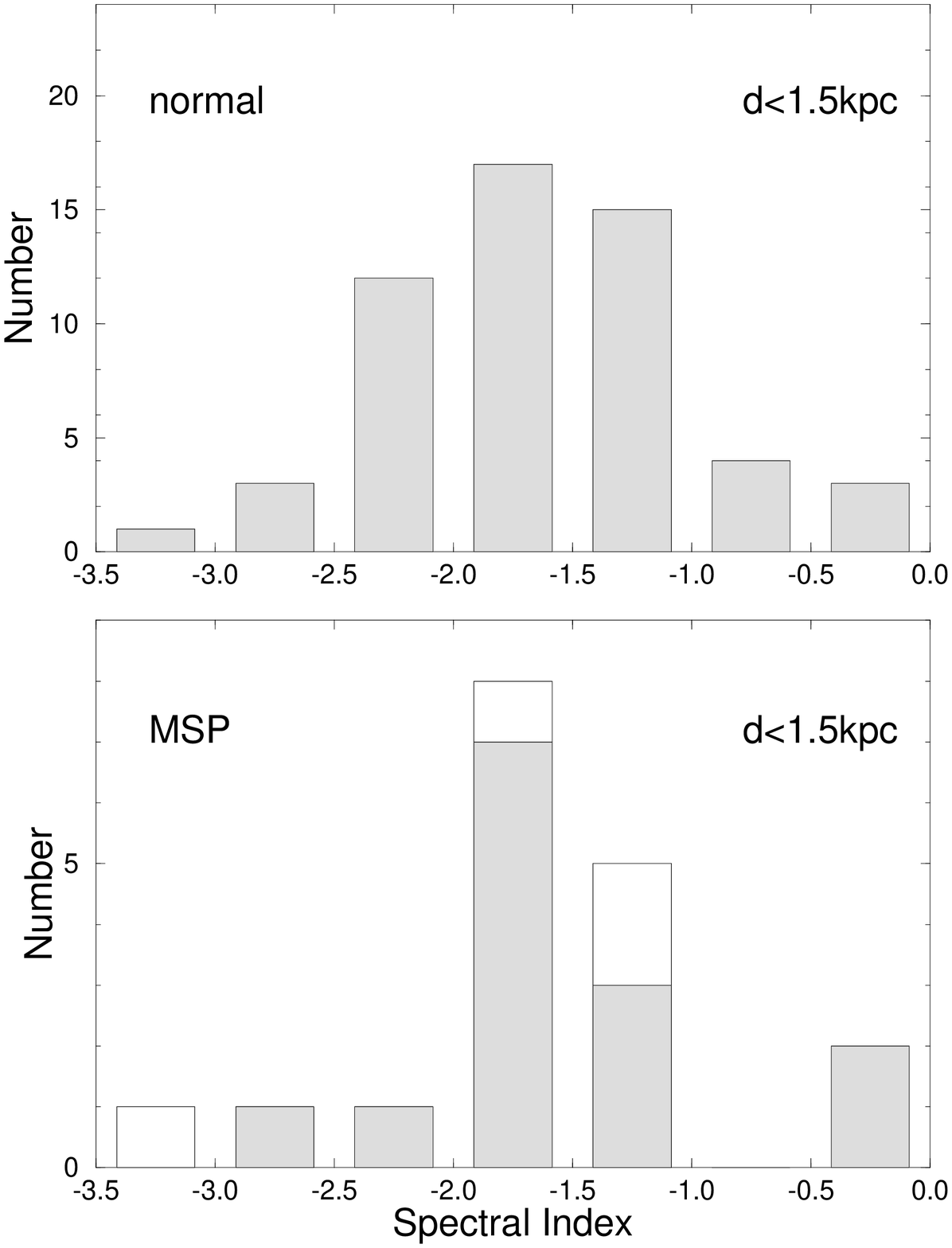}
\bigskip

\figcaption{\label{specindHist2} Distribution of spectral
indices for normal and millisecond pulsars within a 
distance of 1.5 kpc. Data 
for MSPs indicated by unshaded bars are derived from the literature
(see text for details).}
\end{figure}

However, as already discussed, the sample of normal pulsars  is obviously 
biased towards distant, flat spectrum sources.
Removing this bias by restricting the compared data sets to those 
sources which are closer than 1.5 kpc, we actually find mean spectral 
indices which are essentially the same, i.e.~$-1.6\pm0.2$ (MSPs) and 
$-1.7\pm 0.1$ (normal pulsars) with
median values of $-1.65$ and $-1.66$, respectively.
Comparing these distributions of 18 MSPs and 55 normal pulsars
(Fig.~\ref{specindHist2}), we find an even higher 
KS-probability of 50\% that the underlying 
parent distribution is the same. As a final confirmation, we compare
the spectral indices of nearby MSPs and normal pulsars detected
by the Parkes survey. In this case,
we obtain mean spectral indices of $-1.8\pm0.2$ (MSPs) and $-1.7\pm0.1$ 
(normal pulsars) and a KS-probability of 53\% that both distributions
are the same. 
We thus conclude that there is no 
apparent difference in the spectra of normal 
pulsars and MSPs. This is somewhat contrary to the first spectral study of
4 MSPs (Foster et al.~1991) in which MSP spectra appeared to be steeper on
average than normal pulsars. The larger sample of objects available to us 
shows that steep spectra MSPs are the exception rather than the rule.

Obtaining more data at high and low frequencies in the 
future, i.e.,~above 2 GHz and below 400 MHz,
will provide interesting information, particularly
if MSP spectra also show the typical steepening at a few GHz and a low frequency
turn-over, which are both often observed for normal pulsars
(Malofeev et al.~1994). Indeed, the spectrum of B1937+21 presented
by Foster et al.~(1991)  indicates a low-frequency turn-over.

\placefigure{specindHist2}

\subsection{Luminosities of MSPs}

\begin{figure}
\epsscale{0.5}
\plotone{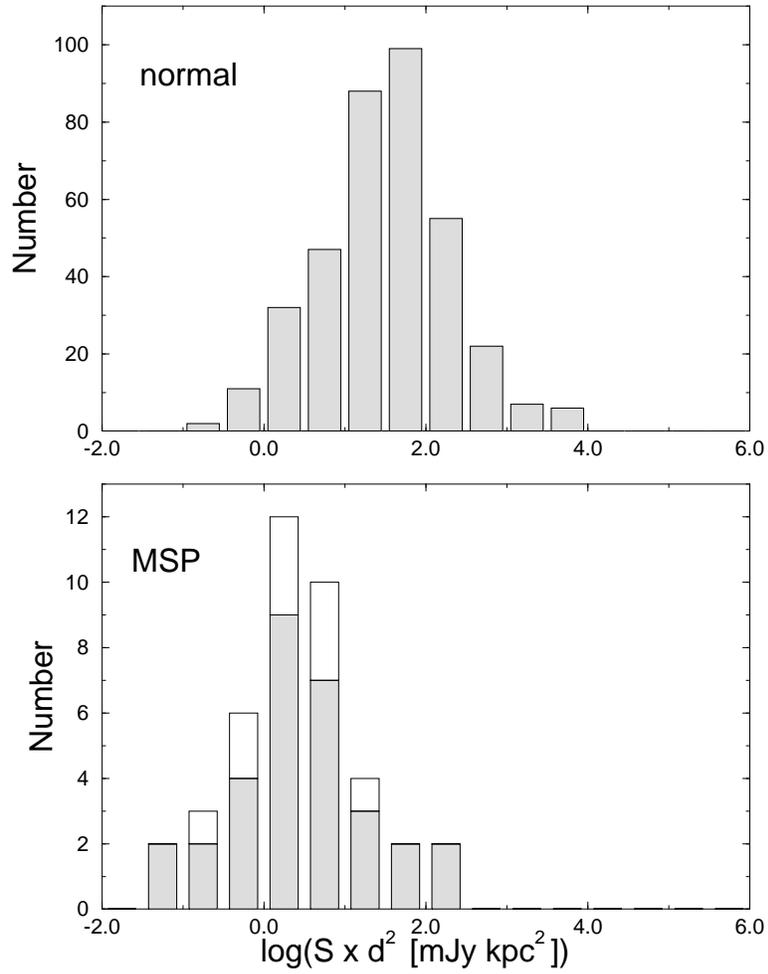}
\bigskip

\figcaption{\label{lum1400} Luminosity distribution derived for normal
and millisecond pulsars. Data 
for MSPs indicated by unshaded bars are taken from the literature
(see text for details).}
\end{figure}

For birth-rate studies of MSPs it is particularly important to compare 
their luminosities to those of normal pulsars. 
%
Since we can estimate the uncertainties of our homogeneous sample
of flux densities obtained at 1.4/1.7 GHz
better than those of values published for 400 MHz by various authors, 
we compare 
distance-corrected flux densities observed at 1.4 GHz for both MSPs and
normal pulsars.
This is possible since we have seen that the spectra of MSPs and normal pulsars
are essentially the same. Consequently, 
we calculate values $S\times d^2$,
i.e. flux density measured in mJy (Table~\ref{fluxtab}) 
times square of the distance in kpc$^2$.
We find that MSPs are apparently one order of magnitude 
less luminous than normal pulsars (see Fig.~\ref{lum1400}). The mean value of
$\log( S \times d^2 \left[ \mbox{\rm mJy kpc}^2 \right])$ 
is only $0.5\pm 0.2$ for our sample of 31 MSPs located in the 
Galactic disk\footnote{We did not include PSR J0218+4232 because for this
object the distance information is very uncertain (Navarro et al.~1995).}
compared to $1.50\pm 0.04$ for 369
normal pulsars (Lorimer et al.~1995b; Taylor et al.~1995). 
This is the reason why MSPs appear
as relatively weak sources at high radio frequencies, so that 
only a small number
of them could be detected at frequencies as high as 4.85 GHz (Kijak et
al.~1997a). Removing the obvious bias from both samples
by restricting our analysis to all those sources which are closer
than 1.5 kpc, we note that this difference in luminosity becomes less
prominent (Fig.~\ref{lum1400.lt.1.5}).
We now find mean values for $\log (S\times d^2)$ of $0.0 \pm 0.1$
(18 MSPs) and $0.57\pm 0.09$ (55 normal pulsars). The corresponding medians
are 0.1 (MSP) and 0.5 (normal), respectively.
\placefigure{lum1400}
Performing a Kolmogorov-Smirnov test on both samples, we find that 
both distributions are similar with a probability of 13\%. 
It is interesting that besides a deviating mean value for the luminosity 
there seems to be a lack of highly luminous sources, 
i.e.~$\log(S\times d^2)>1$,
found for MSPs relative to normal pulsars. This 
could be an artifact due to the smaller number of MSPs in our sample
or perhaps due to a possible remaining small difference in the spectral
distribution of both samples. However, relying completely on published
flux densities we find a similar trend in the 
(usually) quoted 400-MHz
luminosities. It therefore seems that, although normal pulsars and MSPs
have fairly similar luminosity distributions, MSPs tend to be slightly
weaker sources on the average. In order to verify this result
we have performed the same analysis on nearby sources detected in
the Parkes survey. The derived values are
$-0.3\pm0.3$ and $0.5\pm0.1$ for the mean logarithm of $S\times d^2$
of MSPs and normal pulsars, while we find 
medians of 0.1 (MSP) and 0.6 (normal), respectively.
A KS-probability of only 4.8\% is found that both distributions  are 
drawn from the same parent distribution.
Summarising, we have indications that the luminosity 
distributions of both types of pulsars are slightly different.
%
\placefigure{lum1400.lt.1.5}

\subsection{Relationships to other parameters}

 Any difference in the emission properties of MSPs
and normal pulsars could have its origin in the assumed special
evolutionary history of MSPs or the significant difference in period
and magnetic field (see also paper II).

It is therefore interesting that although we have only seven 
(out of nine) isolated 
MSPs (including the planetary system B1257+12 ) in our flux density sample, 
we can confirm the observation
by Bailes et al.~(1997), that isolated MSPs tend to exhibit lower
luminosities than binary MSPs. The mean value of 
$\log(S\times d^2)$ for isolated MSPs is only $-0.1\pm0.4$ and thus
lower than that for the rest of the sample, $0.6\pm 0.1$. 
Restricting the sample to those sources within a distance 
of 1.5 kpc, we still find a difference in the mean value for isolated
MSPs ($-0.5\pm0.3$) and binary MSPs ($0.2 \pm 0.1$). 
In contrast, the mean spectral index for isolated MSPs appears to be
the same as that for the whole sample, $-1.7\pm0.1$.

Investigating possible relationships of the spectral properties of MSPs
to spin parameters like period, magnetic field, spin down luminosity
($\propto \dot{P} P^{-3}$) 
or accelerating potential
($\propto B/P^2$, e.g.,~Ruderman \& Sutherland 1975), we could not
find significant correlations. Calculating these values we have used 
intrinsic period derivatives, i.e.,~observed values of $\dot{P}$ were 
corrected for kinematic effects whose importance for pulsars was first
pointed out by Shklovskii (1970). As a result,
measured spin-down rates can be biased by
proper motion and Galactic acceleration (see Damour \&
Taylor 1991), which can become a significant effect for 
MSPs (Camilo, Thorsett, \& Kulkarni~1994).
For those sources where no
velocity information was available, after Camilo et al.~(1994) we
assumed a transverse speed of 75 km/s.  
We note in passing, that 
there might be a tendency for the spectral index of MSPs to
correlate with the characteristic age of MSPs, 
i.e.,~younger MSPs tend to have a steeper spectrum than older MSPs.
We find that Galactic disk
MSPs with a characteristic age between $10^8$ yr and
$10^9$ yr have a mean spectral index of $-2.5\pm0.3$, while those with a
characteristic age between $10^9$ yr and $10^{10}$ yr exhibit
a mean spectral index $-1.9\pm0.2$. MSPs with a characteristic age 
larger than $10^{10}$ yr in comparison show a mean index  of $-1.4\pm0.2$.
This trend would be in apparent contrast to that observed for
normal pulsars (Johnston et al.~1992; Lorimer et al.~1995b). 
As new sources are discovered, it will be interesting to see whether
this tendency still holds.
However, the lack of any significant correlation between spectral properties
and observed spin parameters is similar to the observations made
for normal pulsars assuming a canonical pulsar model (Malofeev 1996).

\begin{figure}
\epsscale{0.5}
\plotone{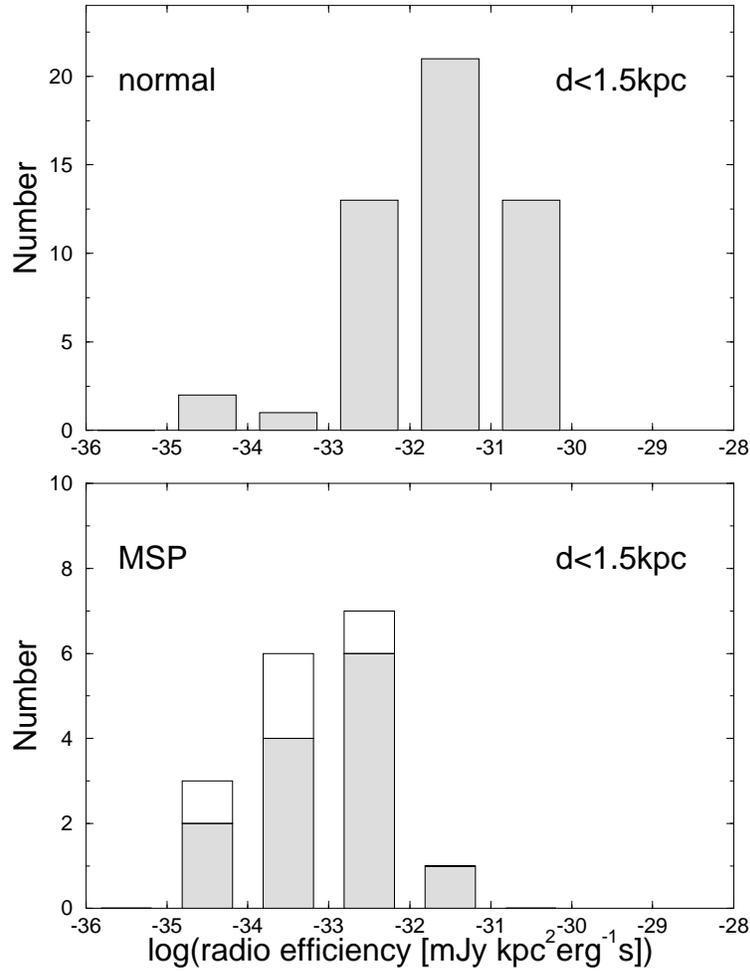}
\bigskip

\figcaption{\label{efficiency} Comparison of the efficiency of normal
and millisecond pulsars as radio sources. Only those objects
located within a distance of 1.5 kpc are considered. Data 
for MSPs indicated by unshaded bars are derived from the literature
(see text for details).}
\end{figure}

Investigating the radio efficiency of MSPs and normal pulsars we have
compared the ratio of spin down luminosity, 
$\dot{E} = 3.95\cdot 10^{46} \dot{P} P^{-3}$ erg/s where $P$ is
measured in seconds 
(e.g.,~Manchester 
\& Taylor 1977), and observed radio luminosity. The distribution
of the logarithm of this ratio, $S\times d^2/\dot{E}$,
expressed in units of mJy kpc$^2$ erg$^{-1}$ s
is presented in Fig.~\ref{efficiency} for both MSPs and normal pulsars.
Again we have restricted  both samples to sources closer than 1.5 kpc
to remove the bias previously discussed. 
It is evident that MSPs are
much less efficient radio pulsars than slow rotating ones. The
mean value of $\log (S\times d^2/\dot{E})$ found for MSPs and normal
pulsars is $-33.3\pm0.2$ and $-31.5\pm0.1$, respectively, while
the mean value of 
 $\log \dot{E}$ itself is found to be $33.2\pm0.1$ (MSP) and
$32.0\pm0.1$ (normal), respectively. 
\label{effval}

\placefigure{efficiency}

\section{Pulse shapes of millisecond pulsars}
\label{beam}

\begin{figure}
\epsscale{0.8}
\plotone{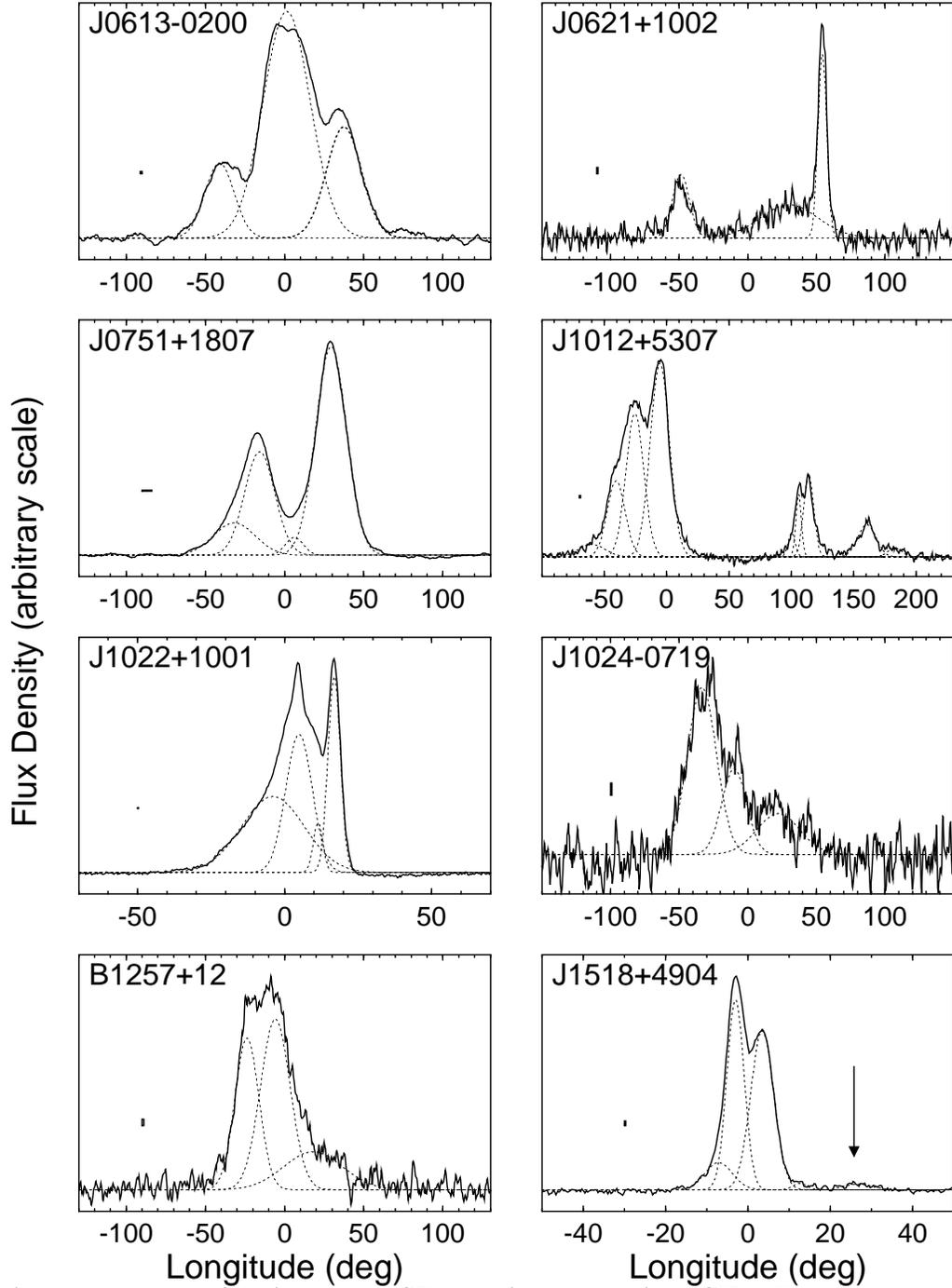}
\bigskip

\figcaption{\label{profa} Average pulse profiles for eight MSPs at a
frequency of 1.4 GHz. The root mean square (RMS) measured in an off-pulse
region is indicated by the height of the small box left of the pulse. Its width
indicates the resolution of the profile (see text for details).
Residuals obtained from fitting Gaussian components (dashed lines) to the
profiles are not plotted for clarity.
The arrow shown in the profile of J1518+4904 points toward a detected post-cursor.}
\end{figure}

\begin{figure}
\epsscale{0.8}
\plotone{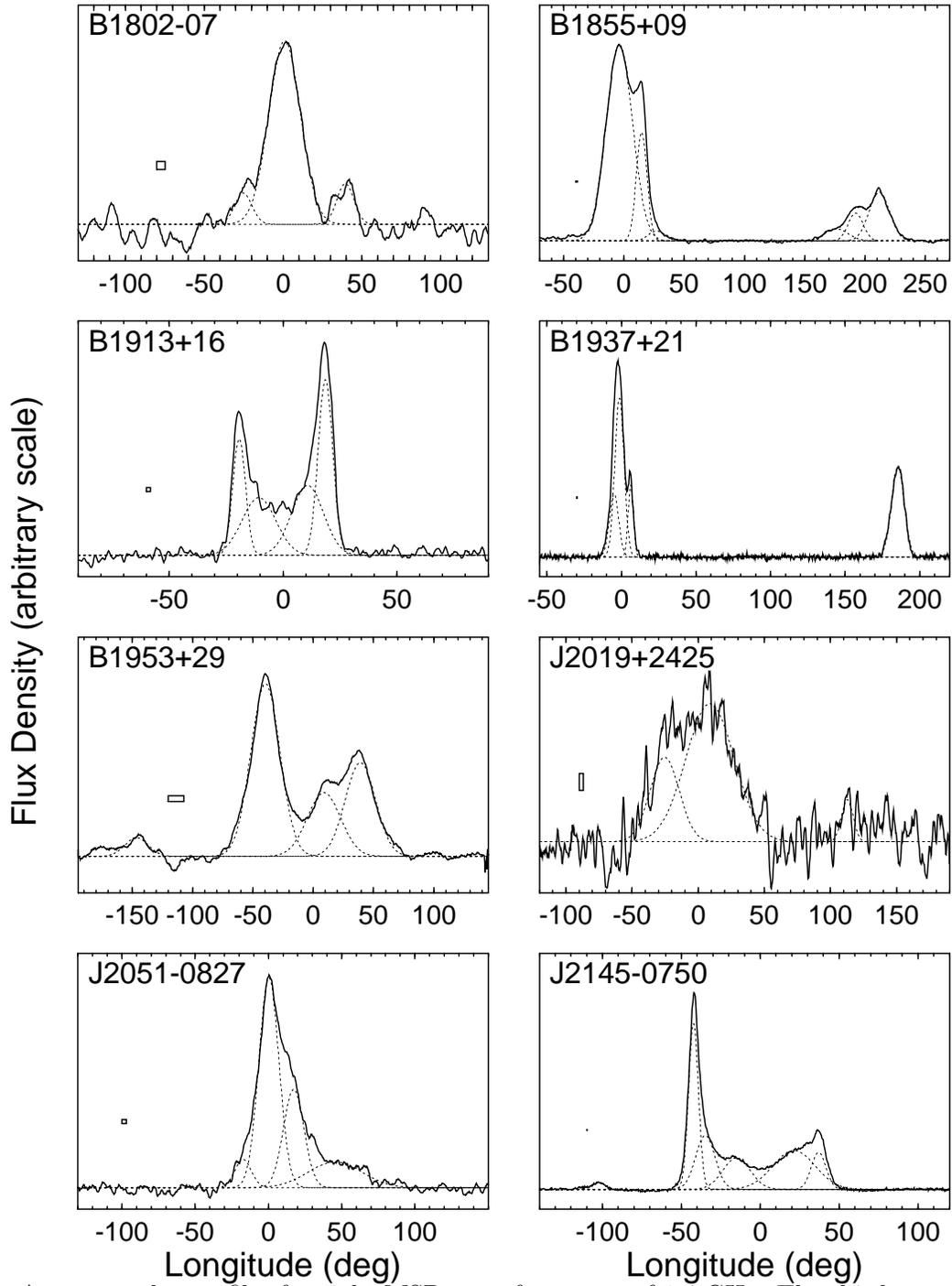}
\bigskip

\figcaption{\label{profc} Average pulse profiles for eight MSPs at a
frequency of 1.4 GHz. The third component tentatively
fitted to the low signal-to-noise-ratio profile of J2019+2425 
at about 104\degr{} corresponds to component $D$ seen by Nice(1992).}
\end{figure}

\begin{figure}
\epsscale{0.8}
\plotone{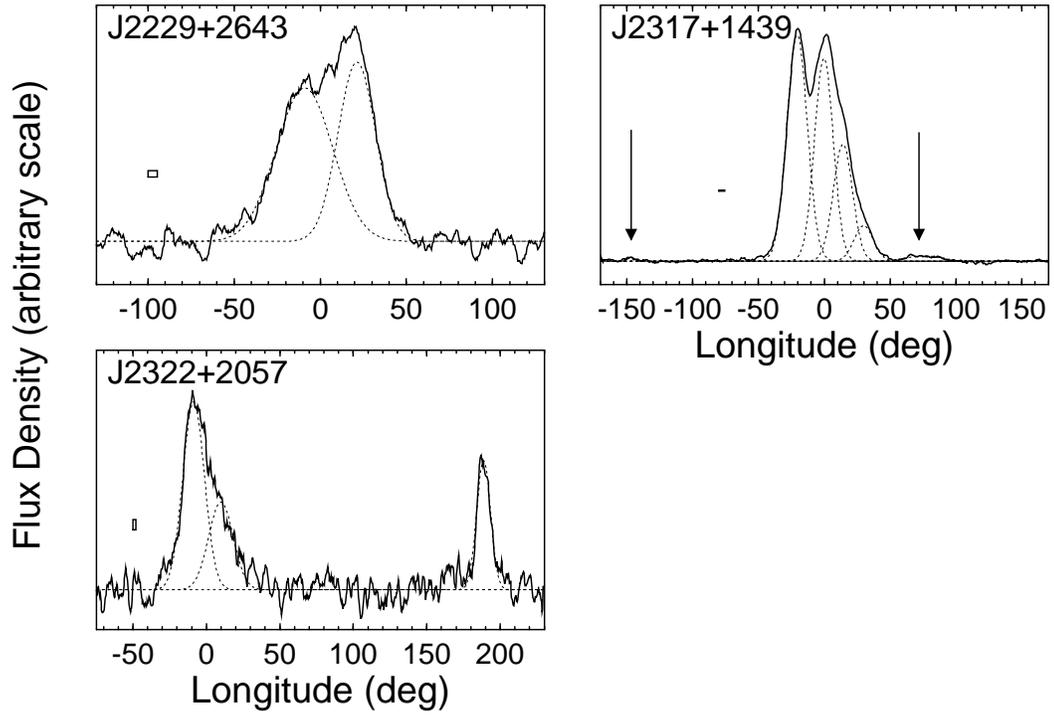}
\bigskip

\figcaption{\label{profd} Average pulse profiles for eight MSPs at a
frequency of 1.4 GHz. 
The arrows shown for J2317+1439 point toward a detected pre- and post-cursor 
of the pulse.}
\end{figure}

In Figs.~\ref{profa}--\ref{profd} we present a collection of 27 profiles of MSPs. 
\footnote{These data are freely available
in the European Pulsar Network Data Archive: 
{\it http://www.mpifr-bonn.mpg.de/pulsar/data/}}
They all represent averages of typically ten to twenty
independent observations and were used as templates
for timing purposes. Most of the profiles 
are of unprecedented quality in signal-to-noise ratio and 
resolution. 
Both are indicated by the small error box left of the pulse.
The height of the box corresponds to one standard deviation (RMS) measured
in an off-pulse region while its width represents the resolution of the
profile. The latter is determined by the dispersion smearing across
one filterbank channel of 666 kHz. 
Only for B1744$-$24A and B1937+21 is this
dispersion smearing apparently large enough to smear out individual features of
the pulse shape, also having a significant impact on the measured width
of the profile presented in Table~\ref{widths}. For B1937+21 we therefore 
present a pulse profile which was obtained with the EBPP, a coherent 
dispersion removal processor recently installed in Effelsberg. 
Details of this system are published elsewhere (Backer et al.~1997).
For the other profiles presented the indicated effective 
resolution of the profile, also quoted in Table~\ref{widths}, 
is that imposed by interstellar dispersion, since we sampled each 
profile with a 
resolution of $P/1024$ s which is generally smaller than the smearing time. 
In the few cases, where sampling time and dispersion
smearing become comparable, the effective resolution is the sum in 
quadrature of both.

\subsection{Newly detected morphological features}

A detailed discussion of individual pulse profiles is presented in
paper II using their polarisation properties to gain further
insight in the nature of the various components. In this paper we want
to draw attention to a few particular sources
where the quality of the profiles obtained is so good, that previously
undetected profile features have been discovered  or where previously
known features are significantly better resolved in our observations.

\begin{figure}
\epsscale{0.8}
\plotone{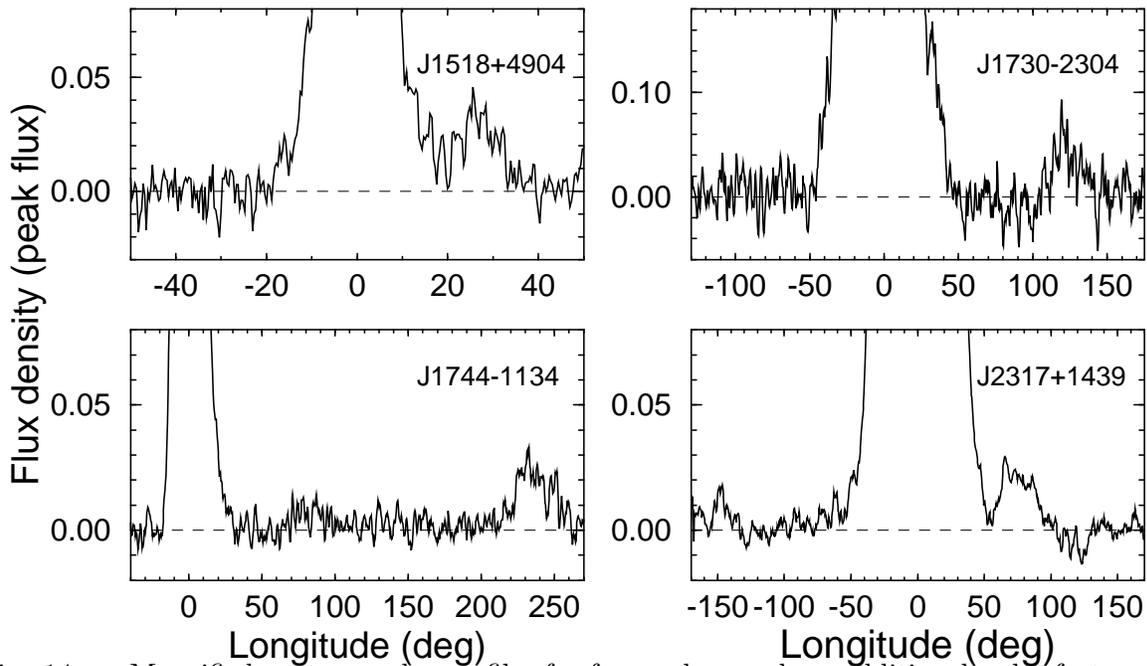}
\bigskip

\figcaption{\label{blownup} Magnified average pulse profiles for four pulsars, where
additional pulse features have been detected.}
\end{figure}

As a first example we note J1518+4904 whose pulse is relatively narrow
compared to the majority of MSPs.
It exhibits a significant post-cursor (see arrow 
in Fig.~\ref{profa}). This post-cursor 
follows the midpoint of the main profile after about 25\degr{} and seems
to be connected to it by  low intensity emission. Details can be
better seen in Fig.~\ref{blownup} where we plot the bottom of the profile
magnified.

\begin{figure}
\epsscale{0.8}
\plotone{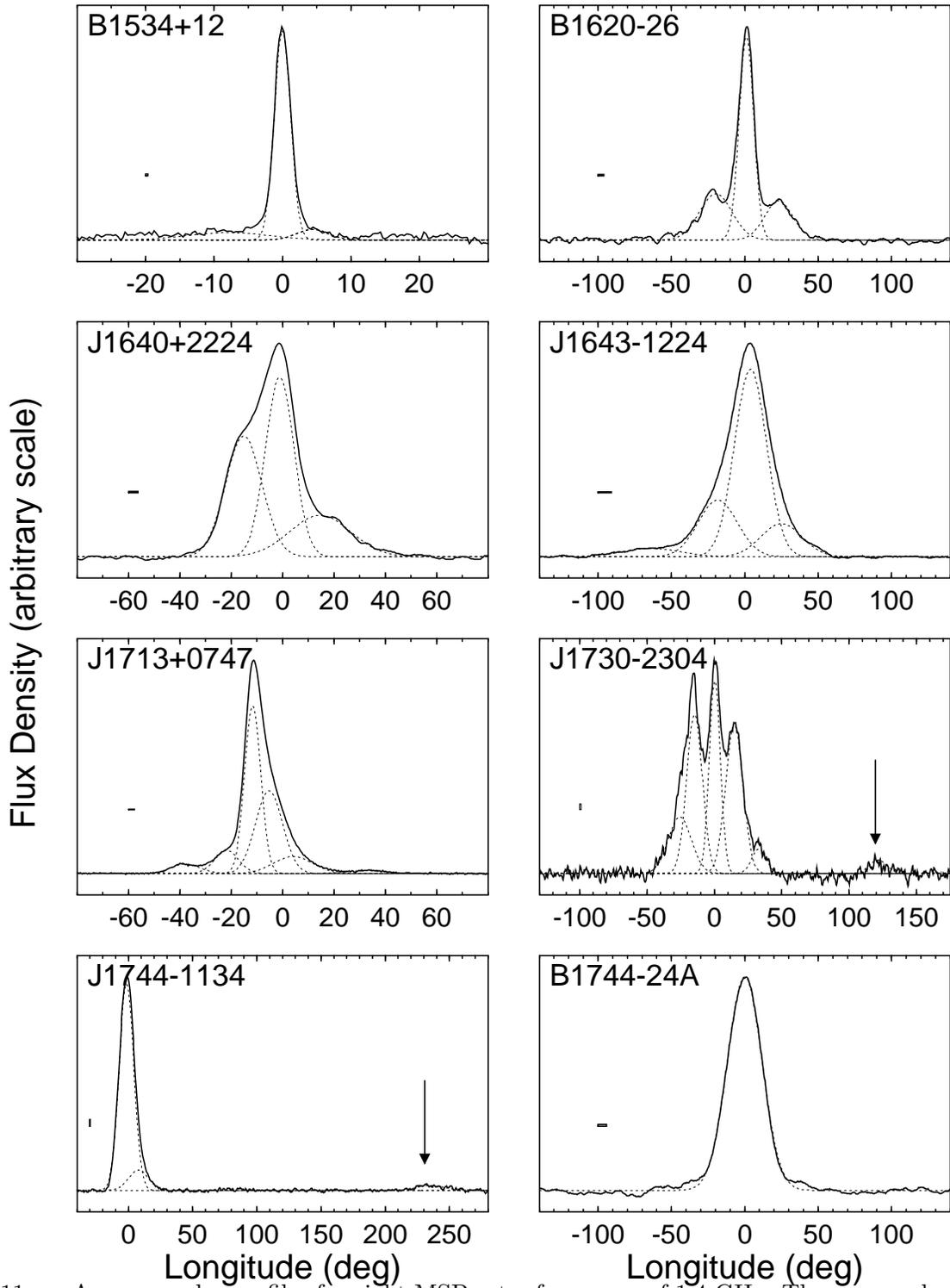}
\bigskip

\figcaption{\label{profb} Average pulse profiles for eight MSPs at a
frequency of 1.4 GHz. 
The arrows shown for J1730$-$2304 and J1744$-$1134 point toward detected 
post-cursor and pre-cursor, respectively.}
\end{figure}

Another post-cursor is detected for J1730$-$2304, which is separated
from the midpoint of the profile by about 120\degr{} 
(cf.~arrow in Fig.~\ref{profb}). We could not detect
any emission connecting main pulse and post-cursor (see 
Fig.~\ref{blownup}).

Another interesting case is J1744$-$1134 which exhibits a pre-cursor
preceding the main pulse by about 124\degr{} (or following the main pulse
by about 236\degr{} as in Fig.~\ref{profb}).
Inspecting the magnified profile in
Fig.~\ref{blownup} we also note the possible existence of a
feature following the profile by about 80\degr. Further observations are
required to establish the existence of this post-cursor.

While the existence of the pre-cursors of J2145$-$0750 
is well known (Lorimer 1994), we can confirm the pre-cursor of B1953+29 first
observed by Thorsett \& Stinebring (1990, 
cf.~Fig.~\ref{profc}). We finally draw the attention to J2317+1439.
For this pulsar we detect a prominent post-cursor, following the
midpoint of the profile by about 70\degr{} but connected to the main
profile by low level emission. Additionally, we note the possible
existence of a weak interpulse preceding the main pulse by about
150\degr.

\subsection{Complexity of pulse profiles}
\label{components}

The complicated 
profiles of some of the first discovered MSPs (e.g.,~B1953+29, B1957+20) 
created the impression that MSP profiles are much more complex than those
of normal pulsars which would be tempting to relate to the evolution 
history of MSPs by a possible disturbance
of the magnetic field structure due to mass accretion (Ruderman 1991).
As now about 40 MSPs  are known in the Galactic plane,
it seems adequate to re-examine this issue again in the
light of a much larger sample presented here.

A measure of the complexity of pulse profiles is obviously given by 
the number of Gaussian components needed to obtain a representation of the 
pulse profile. Using a method described by
Kramer et al.~(1994),
which assures that random noise features are not misinterpreted
as spurious pulse components, we follow Foster et al.~(1991) and
have separated the pulse shapes into individual Gaussian components.
For each profile, the components obtained are indicated by the dashed lines in
Fig.~\ref{profa}--\ref{profd}. 
In order to compare them to a large sample of normal pulsars,
we applied the same component separation method to 180 profiles of normal
pulsars presented by Seiradakis et al.~(1995).
Since all profiles
were observed at the same frequency using the identical observing
system EPOS and by applying the same time
resolution, both samples are ideal for a comparison
of the profiles of normal pulsars and MSPs.
Additionally, due to the very interesting shape of PSR J0437$-$4715,
we included the 1520 MHz profile 
presented by Bell et al.~(1997) in our analysis
which is comparable in quality and resolution to the Effelsberg data.

\begin{figure}
\epsscale{0.5}
\plotone{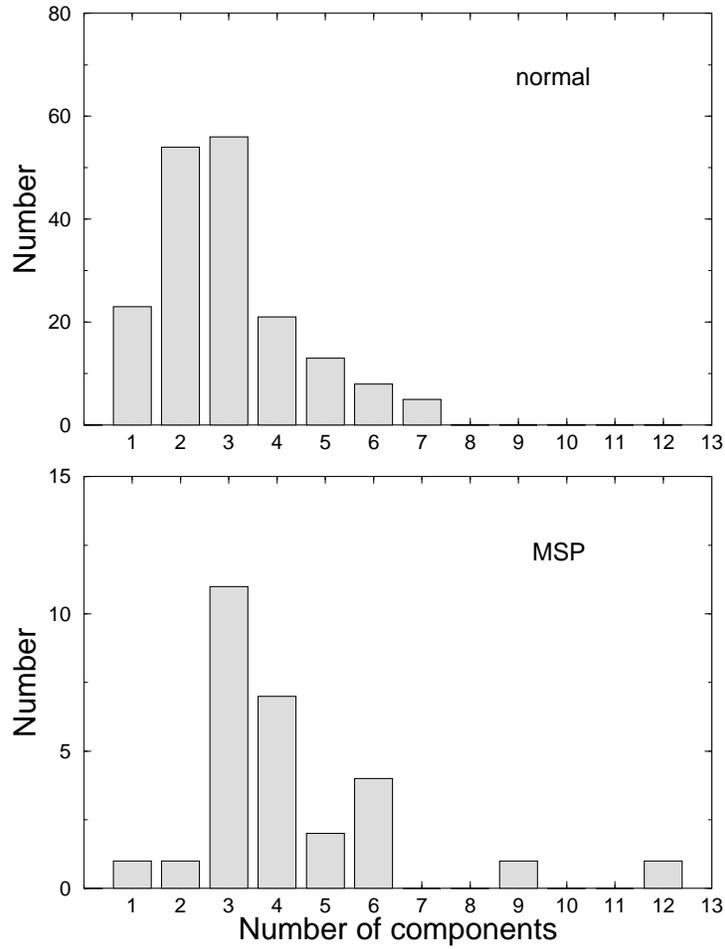}
\bigskip

\figcaption{\label{comphist} Complexity of profiles for normal 
and millisecond pulsars as expressed by number of components in each profile.}
\end{figure}

\placefigure{comphist}

Separating MSP profiles into components we find that
they exhibit on average $4.2\pm0.4$ Gaussian components. For normal pulsars
we derive a mean number of components, which is smaller by one
unit, $3.0\pm0.1$. The sample of normal pulsars contains a large
number of profiles with more than five components. This is mostly due
to the existence of multi-component interpulses (e.g.,~B1929+10) but
also due to such complex profiles as B0740$-$28 (cf.~Kramer 1996)
or B1237+25, B1742$-$30 and B1952+29 (Seiradakis et al.~1995).
On the other hand, the sample of MSPs contains  the very
complex profiles of J0437$-$4715 (12 components)
and J1012+5307 (9 components). Comparing the medians
of the number of components, we 
however, find the same result, i.e.,~a median of 4 for MSPs compared to 3
for normal pulsars.
Fig. \ref{comphist} shows the distribution of both samples,
demonstrating that the majority of MSP profiles can be described by
only three to four components. 

It is important to note that in
the analysis described we included components representing interpulses or post- and
pre-cursors. While only about 2\% of all normal pulsars
are known to exhibit interpulses or pre/post-cursors 
(Lyne \& Manchester 1988; Seiradakis et al.~1995; Taylor et al.~1995), about 36\% of
all Galactic MSPs known either emit a post- or pre-cursor, or exhibit
an interpulse. This apparent characteristic of MSPs has been noted before
in particular for the occurrence of interpulses,
albeit in a much smaller sample (Ruderman 1991). 
This is an important distinction between these two classes of objects,
which also accounts for some of
the difference in the complexity.
In summary, we conclude that the difference 
in the  complexity of pulse shapes of MSPs and normal pulsars
is surprisingly small (i.e.,~only one Gaussian component). Simultaneously,
we consider the peculiar large number of MSP interpulses and pre/post-cursor
as an important clue
in deciphering the circumstances in their very compact magnetospheres 
(cf.~paper II).

\subsection{Shape of the emission beams}

If the emission beam of radio pulsars is confined to the open field
line region, simple scaling arguments based on the opening angle of the
last open field lines (Goldreich \& Julian 1969) 
suggest that the radiation beam of
MSPs should be larger than that of normal pulsars. As a first
approximation for the size of the beam one can study the observed
profile widths which are listed in Table~\ref{widths} for our sample
of MSPs. These widths were obtained using the technique described by
Kramer et al.~(1994) and are quoted for a 50\% and 10\% intensity
level, which refer to the peak of the outermost resolved components.
Quoted 10\%- and 50\%-widths, $W_{\rm 10}$ and $W_{\rm 50}$, 
represent only the value measured for the main pulse, 
i.e.,~possible interpulses are not included. Post- and pre-cursors
are only considered in the calculations if their intensity exceeds the
10\% intensity level. Additionally, we present the equivalent
pulse width, $W_{\rm eq}$, defined as the width of a box-car like pulse shape of the 
same energy and amplitude as measured for the 
real pulse. Generally, a large
ratio of $W_{\rm 50}/W_{\rm eq}$ indicates the presence of prominent outer
components. All widths and their corresponding uncertainty, which is
mainly determined by the dispersion smearing, are quoted in both units of
degrees of longitude and milliseconds. Additionally, we used the value of
$W_{\rm 10}$ to derive the duty cycle of the pulsed emission.

\begin{figure}
\epsscale{0.5}
\plotone{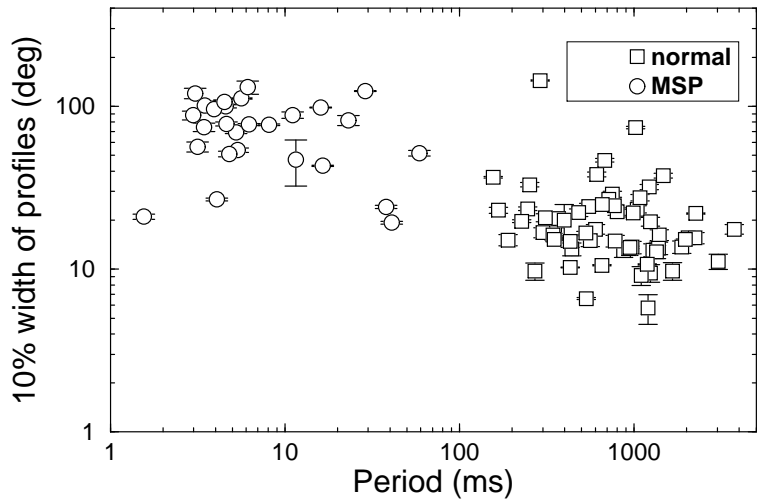}
\bigskip

\figcaption{\label{widthfig} Profile width measured at a 10\% intensity
level for normal pulsars (squares) and MSPs (circles) 
as a function of pulse period.}
\end{figure}

MSP profiles typically have a considerably larger duty cycle
than normal pulsars, as indicated
by Fig.~\ref{widthfig}, where we have plotted $W_{\rm 10}$ as a function of
pulse period. Values of $W_{\rm 10}$ for normal pulsars were taken from
Gil, Kijak, \& Seiradakis~(1993) and Kramer et al.~(1994) who used the same
observing system and frequency as used to obtain the data
presented here. 
Although the effects of dispersion and sampling  
generally increase the width
of the MSPs more than the normal pulsars, the {\em relative}
resolution of our MSP data (i.e.~4.2\% of the period 
in the worst case of B1744$-$24A compared to more typical values 
of less than 1\%) is very similar to that of normal pulsars,
so that the generally larger duty cycle is clearly significant.
However, the pulse width is only a poor estimator
for the true size of the emission beam since it is significantly biased
by the individual viewing geometry of each  pulsar. 
\placefigure{widthfig}

The observed pulse
width depends strongly on the inclination angle $\alpha$ between the
rotation and magnetic axis and the impact parameter $\beta$, which
describes the closest (angular) approach of our line-of-sight to the
magnetic axis. Usually, $\beta$ is defined to be positive if 
line-of-sight and rotation axis are located on opposite sides of the
magnetic pole (cf.~Lyne \& Manchester 1988). In this framework, the
actual angular {\em radius} of the emission beam, $\rho$, and an observed
pulsed width, $W$, are related by
\begin{equation}
\label{rhoeq}
\sin^2 \left(\frac{W}{4} \right) =
\frac{\sin^2 \left( \frac{\rho}{2} \right) - 
       \sin^2\left( \frac{\beta}{2} \right)}{
       \sin  \alpha \cdot \sin(\alpha +\beta) }
\end{equation}
(e.g.,~Gil, Gronkowski, \& Rudnicki~1984). 
Knowing the emission geometry, one can use $\alpha$,
$\beta$ and an observed width $W_{\rm 10}$ to derive the opening angle of
the beam at a 10\%-level, $\rho_{\rm 10}$, which corresponds to the angular 
radius of a circular radiation beam. 

\begin{figure}
\plotfiddle{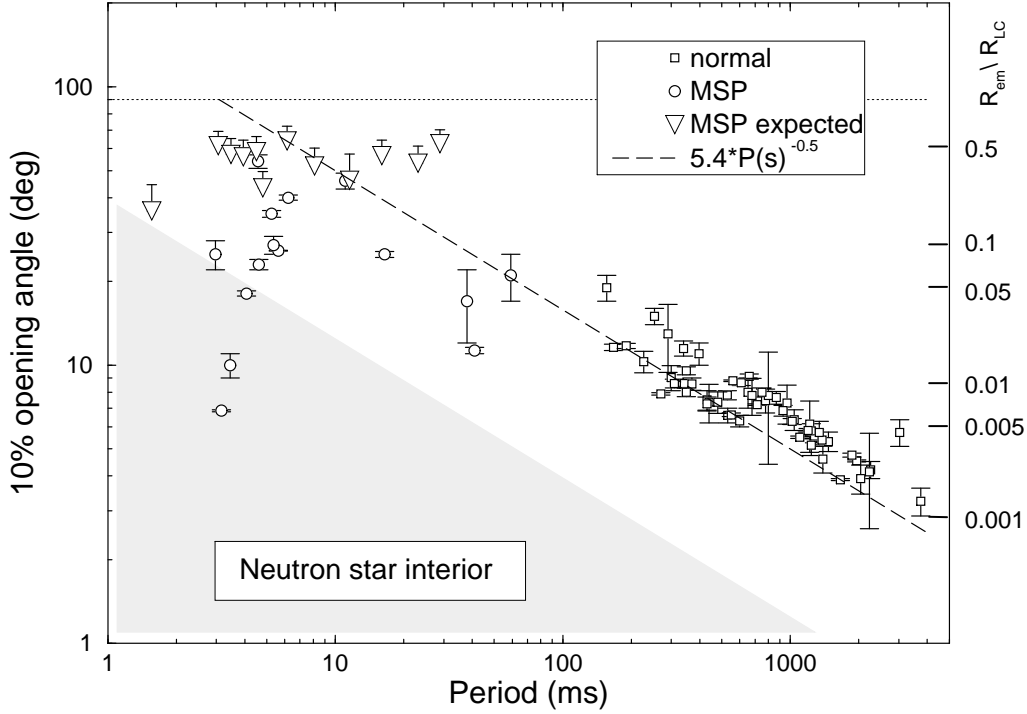}{10cm}{270}{60}{60}{-250}{350}
\bigskip

\figcaption{\label{rhofig} Opening angles derived for a 10\% intensity
level for normal pulsars (squares) and MSPs (circles) as a function 
of pulse period. For normal pulsars, the opening angles scale with
$P^{-0.5}$ as indicated by the dashed line derived by Gould (1994).
MSPs seem not to follow the same scaling, which is supported by the
determined upper limits also presented. The
triangles mark the expectation value, $\langle \rho \rangle$, 
while the upper bar indicates
an upper limit valid with 68\% probability (see text for details). 
Assuming dipolar magnetic field lines, we have marked the region of
opening angles which corresponds to the interior of a neutron star
of 10 km radius. With the same assumption one can translate an opening
angle into an emission height in units of fraction of the light
cylinder. A corresponding scale is indicated on the right border of the
plot. The maximum possible value of the opening angle of 90\degr{} is
marked by a dotted line.}
\end{figure}

In the ideal case, $\alpha$ and
$\beta$ can be derived from polarisation data by fitting the 
classical rotating vector model
(Radhakrishnan \& Cooke 1969) to the observed swing of the position
angle (e.g.,~Blaskiewicz, Cordes, \& Wasserman~1991; 
Hoensbroech \& Xilouris 1997a).
For normal pulsars the viewing geometry has been determined for a large
sample (e.g.,~Lyne \& Manchester 1988; Blaskiewicz et al.~1991;
Rankin 1993a,b), so that $\rho_{\rm 10}$ could be determined. In 
Fig.~\ref{rhofig} we plot opening angles for normal pulsars
derived from profiles observed at a frequency of 1.4 GHz which were 
again taken from data presented by Gil et 
al.~(1993) and Kramer et al.~(1994). Since the profile width is generally 
a function of frequency, we could not include the large sample presented by 
Rankin (1993b), since these data were scaled to a lower frequency of 1 GHz.
The large  scatter visible in the width-period plot of normal
pulsars (Fig.~\ref{widthfig}) disappears and the opening angle of normal pulsars
seems to follow a $P^{-0.5}$-dependence which had
been pointed out by Rankin (1993a) and later 
confirmed by Gil et al.~(1993), Gould (1994) and Kramer et al.~(1994;
see also Biggs 1990).
This period dependence is the same as that for the opening
angle $\rho_{\rm 0}$ of the last open
field line in a dipolar field structure. For a given emission height
$R_{\rm em}$ one obtains the relationship:
\begin{equation}
\label{rho0}
\rho_{\rm 0} = \sqrt{ \frac{ 9 \pi R_{\rm em}}{2Pc}},
\end{equation}
where $\rho_{\rm 0}$ is measured in radians, $c$ is the speed of light (m s$^{-1}$),
and $P$ is the pulse period (s).

\placefigure{rhofig}

\begin{figure}
\epsscale{0.5}
\plotone{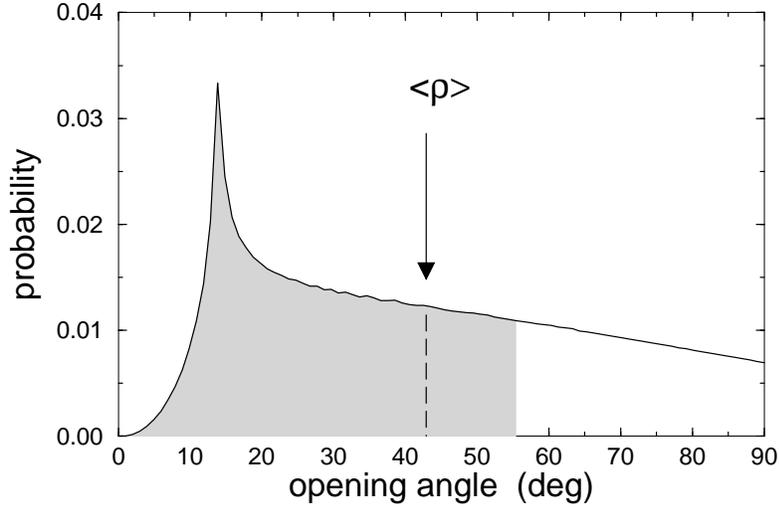}
\bigskip

\figcaption{\label{rhomax} Probability distribution for the occurrence 
of values for the opening angle $\rho$ derived for J1744$-$1134 with 
a 10\%-profile width of 26.8\degr. For this case, the most likely value
of $\rho$ given by the peak of the probability function is 13.8\degr.
The expectation value $\langle \rho \rangle$ is marked
at 42.2\degr. With a probability of 68\% the true value of
$\rho$ is smaller than 55.4\degr{} (shaded area). These values are
to be compared to an opening angle derived from polarisation measurements
of 18.1\degr.}
\end{figure}

While the $P^{-0.5}$-dependence of the opening angle has been
independently determined, the details of the scaling law (e.g., 
the actual scaling factor or a possible bi-modality of the distribution)
differ slightly among the various authors. In any case, the relation obtained 
and the maximum possible value for $\rho$ of 90\degr{}, corresponding to 
a duty cycle of unity and thus continuous emission, limit the period 
to a minimum value possible theoretically for pulsed emission. In other words,
if the scaling law derived for normal pulsars applies also to MSPs, a
detection of MSPs with periods below this critical period might not be
possible. For a frequency of 1.4 GHz, Gould (1994) derives a lower bound  
for the opening angles given by $\rho_{\rm 5} = 5.4^\circ \cdot P^{-0.5}$ 
measured at a 5\%-intensity level, while Gil et al.~(1993) present
$\rho_{\rm 10} = (4.9^\circ \pm 0.5^\circ) \cdot P^{-0.48\pm0.03}$
and Kramer et al.~(1994) 
$\rho_{\rm 10} = (5.3^\circ \pm 0.3^\circ) \cdot P^{-0.45\pm0.04}$
for a lower bound valid at a 10\%-intensity level.
These scaling laws would imply a minimum possible
period between 1 ms and 4 ms, which would be just consistent with the
shortest periods actually observed. Moreover it would suggest that, if
the neutron star equation-of-state allows the existence of
sub-millisecond pulsars, such objects might exhibit substantially different
emission properties. Obviously, it is very interesting to see whether
the scaling laws obviously present for normal pulsars are also valid 
for MSPs. In the following we therefore attempt
to  derive the opening angle of the beam for a number of MSPs.
\placefigure{rhomax}

The polarisation data of MSPs presented in paper II are used 
to model the observed position angle swing in order to obtain information 
about the viewing  geometry and the emission heights (paper III). 
For fourteen sources, fits assuming the
classical rotating-vector-model are made and $\alpha$ and $\beta$ 
determined. Corresponding values derived for the opening angle are presented
in Table~\ref{rhotab} and plotted in Fig.~\ref{rhofig} as a function of
period. The viewing geometry for B1534+12 presented by Arzoumanian et
al.~(1996) and that presented for B1855+09 by Segelstein et al.~(1986)
are consistent with the results of paper III by yielding the
same opening angles within the uncertainties. For PSR B1913+16, Cordes,
Wasserman, \& Blaskiewicz (1990) 
find constraints for the angles $\alpha$ and $\beta$, from 
which we can derive $\rho_{\rm 10}=21\pm4$\degr.

Although fitting the position angle swing is certainly the best
way to obtain estimates of the individual $\alpha$ and 
$\beta$ values (cf.~Lyne \& Manchester 1988; Rankin 1990), 
the results might be sometimes only poorly constrained
(e.g.,~Hoensbroech \& Xilouris 1997a). In fact, the uncertainties of
$\rho_{\rm 10}$ presented in  Table~\ref{rhotab} and Fig.~\ref{rhofig} 
for both normal pulsars and MSPs reflect only the estimated error in
the used pulse width. However, very often the derived opening angles
agree within these uncertainties if values for $\alpha$ and $\beta$
are used which were derived independently by various authors
(see, e.g.,~Kramer et al.~1994). Nevertheless,
we tried to find a second way to get an estimate for the opening
angle which is {\em independent} of derived values for $\alpha$ and $\beta$.
We developed a new method to constrain 
the actual value of $\rho$ statistically. This is described in the
following. 

For an observed width of the
pulse profile, a certain combination of $\alpha$ and $\beta$ (or
$\alpha$ and $\sigma \equiv \alpha + \beta$ as used hereafter) 
leads to a value of $\rho$
fulfilling Eq.~(\ref{rhoeq}). The angles $\alpha$ and $\sigma$ are
both defined in the interval $[0; \pi]$. One can thus test how often
a certain value for $\rho$ could explain  the observed width for the
given parameter space. While we cannot make any {\it ad hoc}
assumption about the distribution of $\alpha$, $\sigma$ as a purely
geometrical factor of the relative orientation of pulsar and Earth,
should be distributed uniformly, so that the observed 
distribution is derived by weighting the possible $\rho$ values by a factor
$\sin \sigma$. For a given profile width we find typical 
probability distributions $p(\rho)$ as presented in Fig.~\ref{rhomax}.
The most likely value for the opening angle is given by the peak of the
distribution function. In the following, we however use the 
typically  larger expectation
value $\langle \rho \rangle$ given
by $\; \langle \rho \rangle \: = \: \int \: p(\rho') \; \rho' \; d\rho'$. 
We can also derive an upper
limit of $\rho$ which is valid with a certain probability. For a 
probability of 68\%, such upper limit $\rho_{\rm max}$ is given by 
$\; \int_0^{\rho_{\rm max}} \: p(\rho') \: d\rho' \: = \: 0.68$. 
Both values are presented
for  a number of MSPs in Table \ref{rhotab} and
Fig.~\ref{rhofig}. The expectation value 
$\langle \rho \rangle$ 
is plotted as a triangle while its error bar indicates the
upper limit $\rho_{\rm max}$, i.e.~with a probability of 68\% the actual
value of $\rho$ is smaller than the one indicated. We stress that 
these statistical estimates do not depend on the determination of 
$\alpha$ and $\beta$.

In Fig.~\ref{rhofig} we compare the opening angles of normal pulsars 
and MSPs. We have plotted the lower bound  $\rho=5.4\cdot P^{-0.5}$
derived by Gould (1994, dashed line). Additionally, we
indicate the maximum possible value of $\rho$ of 90\degr{} by a dotted
line.  Assuming dipolar magnetic field lines (Eq.~\ref{rho0}), 
we have marked the region of
opening angles which corresponds to the interior of a neutron star
of 10 km radius. With the same assumption one can use Eq.~(\ref{rho0})
to translate the opening angle into an emission height in units of 
fraction of the light cylinder radius, $R_{\rm LC}=c \; P /2\pi$. 
A corresponding scale is indicated on the right hand side of the
plot. 

It is obvious from our sample of 27 MSPs that three sources have
statistical values larger than predicted by the scaling law 
derived from the sample of normal
pulsars (J0621+1002, B1802$-$07 and J2145$-$0750), while
only five MSPs
(i.e.~B1620$-$26, J1730$-$2304, B1744$-$24A, B1913+16 and  B1953+29)
exhibit opening angles which are consistent with this scaling law.
In fact, the large majority of opening angles 
derived for MSPs are significantly smaller than expected 
(note the logarithmic scale). A similar statement has been already made
for the pulse width of B1937+21 by Chen \& Ruderman (1993; see
also Backer 1995) although we
have demonstrated that the pulse width itself is only a poor estimator
to learn more about the size of the emission beam. Here we have
derived opening angles and thus attempted to remove geometrical effects
which are always present. For three sources (J1640+2224, J1744$-$1134 and
J2317+1439) the derived values are not only smaller than expected but they
even indicate that the emission takes
place {\em inside} the neutron star. For the other sources, the
indicated emission heights are substantially closer to the neutron star
than those derived for normal pulsars. However, due to the compactness
of MSP magnetospheres, the emission height is nevertheless at a
significant fraction of the light cylinder radius, i.e.~at 10\% to 50\%
compared to the at most few percent observed for normal pulsars (e.g.,~Cordes
1978; Matese \& Whitmire 1980; Cordes \& Stinebring 1984; 
Blaskiewicz et al.~1991; Phillips 1992; Xilouris et al.~1996). 

\subsection{Frequency development of pulse profiles}
\label{freqchange}

Normal pulsar profiles very often show a distinct profile
development with frequency which can be used for a classification
scheme, such as the one devised by Rankin (1983). Comparing the frequency
development of pulse profiles of normal pulsars and MSPs between
400 MHz and 1400 MHz, one gets the impression that MSP
profiles change much less with frequency than normal pulsars.
A detailed discussion of
individual MSP profiles and their frequency development is given in
paper II. Here we just want to investigate whether this tendency
is related to the assumed evolutionary history of MSPs. In such a case, the
profiles could be affected by the amount of mass transferred from the
binary companion since the transferred mass could have altered the dipolar
magnetic field  structure of the progenitor and thus the profile development
compared to normal pulsars (Ruderman 1991). 

In order to quantify the profile development of MSPs we have
used the results of the component separations (see Sect.~\ref{components})
to produce noise-free templates. These templates were modified by
adjusting the individual components to match the profiles observed at
lower frequencies (in cases where low-frequency data were available in 
digital form, these data were used; for references see Table~\ref{fluxtab}). 
The normalised difference in the area defined by the two pulse shapes, 
$\kappa$, was finally used as the parameter
describing the profile change, i.e.~
\begin{equation}
\kappa = \sum_{i=1}^{n} 
\left| \frac{t_i(\nu_1)}{\sum_{j=1}^{n} t_j(\nu_1)}
     - \frac{t_i(\nu_2)}{\sum_{j=1}^{n} t_j(\nu_2)} \right|.
\end{equation}
The noise-free templates obtained for frequencies $\nu_1$ and $\nu_2$,
are represented by $n$ samples, $t_1,...t_n$. 
While $\nu_1$ was 1.41 GHz, $\nu_2$ was
generally taken to be 400 MHz (Table~\ref{fluxtab}). 
A value close to unity means a large
development in frequency while a value close to zero means a small change in
the profiles. The parameter, $\kappa$, could be determined for those
profiles of MSPs in the Galactic disk
for which profiles of considerably good resolution and quality at low and high
frequencies were available. Profiles of B1937+21 were
excluded since they are affected by interstellar scattering at
low frequencies (Thorsett \& Stinebring 1990).

\begin{figure}
\epsscale{0.5}
\plotone{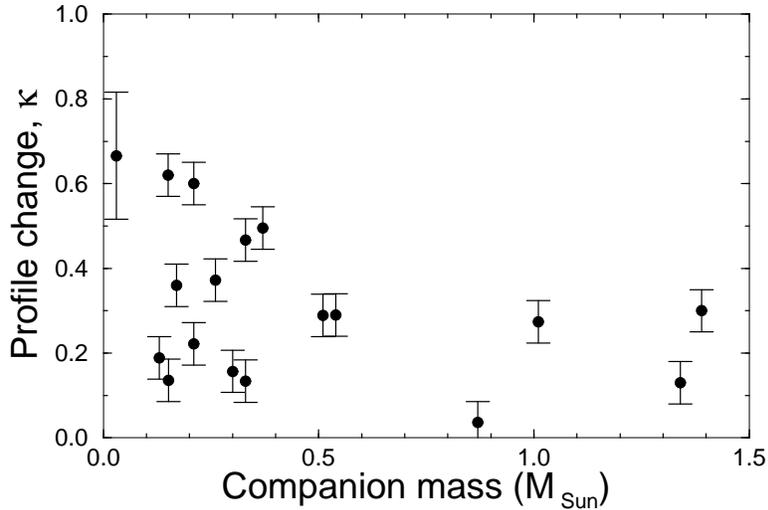}
\bigskip

\figcaption{\label{profevolv} Parameter describing the profile
development between 400 MHz and 1400 MHz (for definition see text) as
a function of companion mass.}
\end{figure}

\placefigure{profevolv}

Since high-mass binary pulsars
are expected to have accreted a smaller amount of mass than low-mass
binary pulsars with large inferred amounts of accreted mass 
(e.g.,~Phinney \& Kulkarni 1994), we plotted $\kappa$ versus
the companion mass (Fig.~\ref{profevolv}). In cases where the inclination
of the binary system is not known, we assumed the most probable 
angle of 60\degr{} to calculate the mass of the companion. 
We observe a 
slight tendency for systems with less massive companions to show 
more profile development with frequency 
than those with more massive companions. 
The mean value of $\kappa$ for systems
with companion masses $m_c\le 0.45M_\odot$ is $0.37\pm0.05$ 
(12 sources) but only 
$0.22\pm0.05$ for those with $m_c>0.45M_\odot$ (6 sources,
Fig.~\ref{profevolv}). 

It can be expected that the final spin period of a recycled pulsar is
related to the companion mass (e.g.,~van den Heuvel \& Bitzaraki 
1995). Studying $\kappa$ as a function of pulse period we see, in fact,
essentially the same tendency. Distinguishing between sources with
spin periods smaller and larger than 10 ms (now also including isolated
MSPs with reasonably resolved profiles), we find a mean $\kappa$ of
$0.35\pm0.05$ for $P\le10$ ms (16 sources) and $0.21\pm0.03$ for 
$P>10$ ms (8 sources). It will be interesting to see whether this apparent
tendency  remains valid as more MSPs
in massive binary systems and with larger spin periods are discovered. 


\section{Discussion}

In Sect.~\ref{fluxes} we have demonstrated that MSPs and normal pulsars
exhibit the same flux density spectrum. While MSPs appear to be 
less luminous compared to normal pulsars, we have also seen
that MSPs are less efficient radio sources. In particular, we 
confirm the observation by Bailes et al.~(1997), that binary MSPs are more
luminous than isolated MSPs, while we show simultaneously that
both groups exhibit the same mean spectral index. Although we observe a weak
tendency that the characteristic age is correlated with the spectral
index of MSPs, no other significant correlation  between spectral
properties of MSPs and other system characteristics is apparent.

Investigating the complexity of pulse profiles of normal pulsars and
MSPs, we have demonstrated that, on average, MSP profiles are only
marginally more complex than those of normal pulsars. In fact, the
intrinsic size of MSP radiation beams is much smaller than
expected from the scaling laws derived for normal pulsars.
We observe the tendency that the profile development with frequency is
related to the amount of mass transferred during the recycling process.

\subsection{MSPs as radio sources}

The sample of sources presented in Table~\ref{fluxtab} misses only two
Galactic MSPs within a distance of 1.5 kpc, which have no published
detection at 1400 MHz (J0034$-$0534 and J1455$-$3330). These
non-detections suggest a steep spectrum for both sources which could
lead to a slightly steeper mean spectral index for all MSPs. The same
however applies also to our sample of normal pulsars where we also
included only those sources which were previously detected at 1400 MHz.
We therefore conclude that the comparison of both samples leads to
reliable results. We note in passing that the mean spectral index of
those four sources detected by Kijak et al.~(1997a) at 4.85 GHz is
$-1.53\pm0.09$ and thus smaller than the mean spectral index 
for the whole sample, as expected.

The result that the spectra of normal pulsars and MSPs are essentially the same, 
strongly suggests the same emission mechanism for both types of
pulsars, in spite of the difference in period by three orders of magnitude.
If the magnetic field structure in the emission region is 
influencing the flux density spectrum, as in the case for curvature
radiation, this result points also towards a dipolar field
structure as found for normal pulsars. In such a case, 
the emission mechanism must also work over
four orders of magnitude in magnetic field strength as inferred by the
measured period derivatives. However, multipole
components existing at the neutron star surface could increase the magnetic
field strength over the value expected from a dipolar field structure
(see Krolik 1991; Arons 1993; paper II). 
In any case, the weak tendency of a correlation between the spectral
index and the characteristic age, which is in contrast to that observed
for normal pulsars, suggests an evolution of the magnetospheric
conditions. 

Our observations show that MSPs tend to be less luminous than normal
pulsars, while they are also less efficient radio emitters compared to
normal pulsars. During the analysis leading to this result we used
a luminosity estimator 
defined as the observed equivalent
continuum flux density, i.e.,~measured pulse energy, $E_{\rm R}$,
averaged over the pulse period, $S = E_{\rm R} / P$, times the 
square of the distance (cf.~Manchester \& Taylor 1977). 
Most of the distances used for both normal pulsars and MSPs are derived
from dispersion measures using the model by Taylor \& Cordes (1993). The 
resulting values are therefore somewhat uncertain. However, our
conclusions are based on samples of objects within 1.5 kpc distance, where
the Taylor \& Cordes (1993) model should be free of systematic trends. 
Any remaining effect not accounted for in this model, would apply to both
groups of objects, and thus would not change our result.
Actually, a more severe effect to be considered is the 
disadvantage of the luminosity estimator in
that it reflects only the emission received by
the cut of our line-of-sight rather than that from the full
emission cone. The derived luminosity is thus also biased by the
viewing geometry. In order to account for this effect one should 
make use
of derived opening angles to use the true size of the emission beam
in order to calculate the luminosity. For a comparison of sources
we had to assume filled emission beams of circular shape. 
Moreover, the opening angles can only be determined for a limited
number of sources, so that
we decided to use a second luminosity estimator based on the peak
flux density, $S_{\rm peak}$, which 
still depends on the viewing geometry but is less affected by
the actual pulse shape and width. 
The previously
used luminosity is defined as $S$(mJy)$\times d^2$(kpc$^2$)$ = 
E_{\rm R}$(mJy s)$\times d^2$(kpc$^2$)$ /P$(s), 
which using the definition of the equivalent width becomes 
$S$(mJy)$\times d^2$(kpc$^2$)
$ = S_{\rm peak}$(mJy)$\times W_{\rm eq}({}^\circ) \times
d^2$(kpc$^2$)$ / 360^\circ$. We have thus computed the value
$L_{\rm peak} \equiv 360^\circ \: S$(mJy)$\times d^2$(kpc$^2$)
$ / W_{\rm eq}$($^\circ$) 
using the equivalent widths
presented in Table~\ref{fluxtab} and those published by Gould (1994) for
normal pulsars.  For normal pulsars within a distance
of 1.5 kpc we find a mean value for
$\log L_{\rm peak}$ at 1400 MHz of $2.11\pm0.09$ with a median of 2.07. 
In contrast, 
for our sample of MSPs closer than 1.5 kpc we find  a mean value for
$\log L_{\rm peak}$ of $1.4\pm0.2$ with a median of 1.5. The KS-probability
that both distributions 
are drawn from
the same parent distribution is only 1\%, strengthening even further
the conclusion that 
MSPs are less luminous than normal pulsars. 


It is interesting
to note  that MSPs seem also to be less efficient gamma-ray
sources since from the six pulsars confirmed as EGRET sources, none is a
MSP (Kanbach et al.~1996). This fact is surprising since the mean value
of $\dot{E}$ for MSPs is more than one order of magnitude larger than
that of normal pulsars (cf.~Sect.~\ref{effval}). Moreover, most of the
known MSPs are nearby sources due to the afore-mentioned selection effects,
so that the ratio of $\dot{E}$ and square of the distance,
$\dot{E}/d^2$, apparently relevant for gamma-ray emission, is much larger
for MSPs than for the average normal pulsar. 
One can speculate whether these observations for radio and gamma-ray
emission are
related since for both types of radiation, similar models are proposed,
i.e.,~polar cap and outer gaps models (cf.~paper II). While the sample
of slowly-rotating pulsars detected as gamma-ray sources is made
up by relatively  energetic pulsars, 
a difference in the particle density in the magnetosphere of normal 
pulsars and MSPs could explain the dissimilarity in their luminosity
distribution. The fact that there is a difference in the 
radio luminosity distribution of binary and isolated MSPs 
strongly suggests that
the luminosity of all MSPs is affected by their evolutionary history. 

\subsection{The beam structure of MSPs}

 It has been suggested that the overall dipolar field apparently
observed for normal pulsars (e.g.,~Phillips \& Wolszczan 1992;
Kramer et al.~1997) could be distorted
by the existence of magnetic multipoles in cases of MSPs
which could be enhanced by mass accretion from a binary companion. 
Disturbance of a dipolar field structure or the existence of 
magnetic multipoles in the emission region should be reflected by
overall shape of pulse profiles, their frequency development, and
in particular in the polarisation properties which are 
discussed in detail in paper II.

\subsubsection{Complexity of pulse profiles}

The previously observed apparent large
complexity of MSP profiles was attributed to a
deviation of the magnetic field structure from a dipolar form
(Krolik 1991). However, we have demonstrated 
that the majority of MSP profiles are actually only 
marginally  more complex. Since most of the profiles used to derive this
result are in general of very high resolution already, we do not expect 
the observable complexity 
to change significantly even if a coherent de-dispersion technique is used
(cf.~profile of B1937+21 and also Backer 1995).

We believe that the impression created after the discovery of only a few
MSPs, was mainly due to the much larger duty
cycle seen (see Table~\ref{widths}). In contrast to the
average 3\% found for normal pulsars (Taylor et al.~1995), we
observe a mean duty cycle of 21\% for MSPs. 
Therefore, for MSPs one immediately observes a ``blown-up'' version of the
pulse shape, enabling an easier identification of various components. 
Zooming in on the profiles of normal pulsars as done by Seiradakis et
al.~(1995) or Kijak et al.~(1997b), can
lead to  similar results for normal pulsars. 
Therefore, the complexity of MSP profiles alone does not obviously  provide
any clue as to 
whether higher order magnetic multipoles exist in the emission
region. In fact, if the number of components is related to the size of
the polar cap, one would  expect a considerably larger number of components
for MSPs even in cases of a dipolar field, since the polar cap scales as
$P^{-0.5}$. This is however not the case. As already discussed, a
perplexing large number of interpulses and pre-/post-cursors
makes up for some of the one unit difference derived in the mean number of 
components of normal pulsars and MSPs. Therefore, even if a dipolar field
near the neutron star or in the emission region exists, the full open
field line region might not be illuminated. In any case, the large number of 
additional pulse features appearing in addition to the main pulse
is an important significant difference 
in the properties of MSP and normal pulsar emission. Combined with
the relatively normal complexity of MSP profiles, it possibly suggests that 
their origin is due to additional
radiation beams, e.g.,~due to outer gap emission (e.g.,~Cheng et al.~1986).
We discuss this possibility further in paper II.

\subsubsection{Beam size}

By investigating the opening angles, we have demonstrated that the radiation 
beams of MSPs are significantly smaller than expected. Certainly, the opening
angles based on a determination of the viewing geometry by studying 
polarisation data (paper III), involves the assumption of the validity 
of the rotating vector model for MSP magnetospheres to derive
$\alpha$ and $\beta$. However, we have seen
that the statistically derived opening angles imply the same result. The
calculation of $\rho$ itself does not include any assumption about the
magnetic field configuration. It is even independent of the actual shape of
the beam, i.e.,~whether it might be circular or elongated in either
latitudinal or longitudinal direction. We suggest three reasons for the
different behaviour of MSPs and normal pulsars. The beams might be 
intrinsically smaller (i.e.,~the open field line region is not completely 
filled by emission in contrast to normal pulsars), the emission height
or a radius-to-frequency mapping might be  different for normal
pulsars and MSPs, or the field configuration in MSPs magnetosphere is indeed
disturbed as compared to a dipolar field presumably dominant in normal pulsars.

Arons (1993) has used the observed spin-down rates to argue that magnetic
multipoles are not prominent in MSP magnetospheres. Again, we focus
on this issue in detail in paper II, where for the first time a large 
homogeneous set of polarisation data is available to trace
the magnetic field structure.  Here we draw 
attention to the fact that by assuming dipolar field lines, the derived
opening angles lead to inconsistent results, i.e.,~emission from {\em inside}
the neutron star. This result is however
partly based on the assumption of the correctness of
the rotating-vector-model fits and a neutron star radius of 10 km. 
As only a ridiculously
small neutron star radius of 1 km can solve this apparent
contradiction, the reliability of the polarisation fits is discussed in
paper III.

For normal pulsars the observed data suggest that the emission height
above the surface is a function of frequency, i.e., high frequency emission
is emitted closer to the neutron star than low frequency emission
(Cordes 1978). If such a model is also valid for MSPs, 
then the observed discrepancy between the 
opening angles of MSPs and normal pulsars could be explained by a
different {\em radius-to-frequency mapping} behaviour, i.e.,~a
different frequency dependence or a change in the absolute scale 
(cf.~Eq.~\ref{rho0}). Comparing the $\rho-P$-relations derived for normal
pulsars at different frequencies (Rankin 1993a,b; Gil et al.~1993; 
Gould 1994; Kramer et al.~1994), 
one notes only a very small  difference in the scaling factor
of the order of one degree of longitude. Given the small
frequency development of MSP profiles (cf.~Sect.~\ref{freqchange})
it becomes obvious that a different frequency scaling cannot account 
for the apparent small beam sizes of MSPs. 
Although the tendency that the profile development is related to 
spin period or the 
amount of mass transferred to the MSP is far from being established,
its confirmation would in fact
suggest that a (probably weak) radius-to-frequency mapping
is acting in a non-dipolar field
structure. Compared to a purely dipolar field, a
small change in the emission altitude should have a
larger impact in a disturbed magnetic field structure, causing a
larger change in the profile with frequency. 
Chen \& Ruderman (1993) argued that mass accretion would
reduce the radius of the polar cap and thus the size of the open field
line region.  Therefore, the recognised
trend in the profile development might not only point towards
the modification of a dipolar field structure by mass accretion, but
could also offer an explanation for the smaller beam sizes.
In any case, it seems  necessary to include additional
independent information to decide whether a multipole or disturbed
dipolar magnetic field or systematically unfilled emission beams are present.
A major step in
this direction is taken in papers II and III, where we focus
on the possible impact of gravitational bending and magnetic field
sweep back. The latter effects are apparently not important for the
emission of normal pulsars (e.g.,~Phillips 1992; Kramer et al.~1997),
but they might become relevant if the emission simultaneously takes
place close to the neutron star surface and at a significant fraction of
the light cylinder radius as indicated by Fig.~\ref{rhofig}.

\subsection{Implications for the birth-rate of millisecond pulsars}

We have seen that normal pulsars and MSPs 
apparently differ in the luminosity distribution and the intrinsic size
of their emission beam. Both properties are important parameters for
deriving the birth-rates of both groups of pulsars. 

In general, one will assume as a first approach the same
luminosity distributions for MSPs and normal pulsars. However,
the observed difference in the mean value of the derived luminosity 
corresponds to a factor of about six and is therefore at a level to become
important for birth-rate studies of MSPs. Indeed, in a recent analysis,
Lyne et al.~(1998) find evidence for a difference in the luminosity function
between normal pulsars and MSPs at low luminosities.

Similarly, one should reconsider the calculated birth-rates of MSPs
under the impression that the scaling law for the beam size derived for
normal pulsars does not apply to MSPs. Assuming a random distribution
of the inclination angle $\alpha$, the beaming fraction, $f$ describing the 
fraction of the sky covered by the radiation beam, is given by
\begin{equation}
f = (1-\cos \rho)+(\frac{\pi}{2}-\rho) \; \sin \rho
\end{equation}
(Emmering \& Chevalier 1989). If the beaming fraction is large, the chance
of detecting a source is high, resulting in a smaller birth-rate 
necessary to sustain the observed population, 
compared to the case of a smaller beaming
fraction. We have seen that if we apply
the scaling law found for normal pulsars also to fast rotating pulsars, the
beaming fraction should be very close to unity, i.e.~$f\approx1$. 
In contrast, inspecting Fig.~\ref{rhofig}, one might get an impression
that $20\degr<\rho<50$\degr{} represents a typical value for MSPs,
corresponding to a beaming fraction of $0.5< f < 0.9$.
An increase in the needed birth-rate can only be avoided if
the radiation beam is not circular in shape, but actually elongated in
latitudinal direction (cf. Chen \& Ruderman 1993). The evidence for 
elongation of the beam is however weak, and while Narayan \& Vivekanand
(1983) have indeed suggested such a beam shape, Biggs (1990) even argued for a
compression of the beam in the
latitudinal direction, depending on the inclination angle.

\section{Summary}

\begin{figure}
\epsscale{0.5}
\plotone{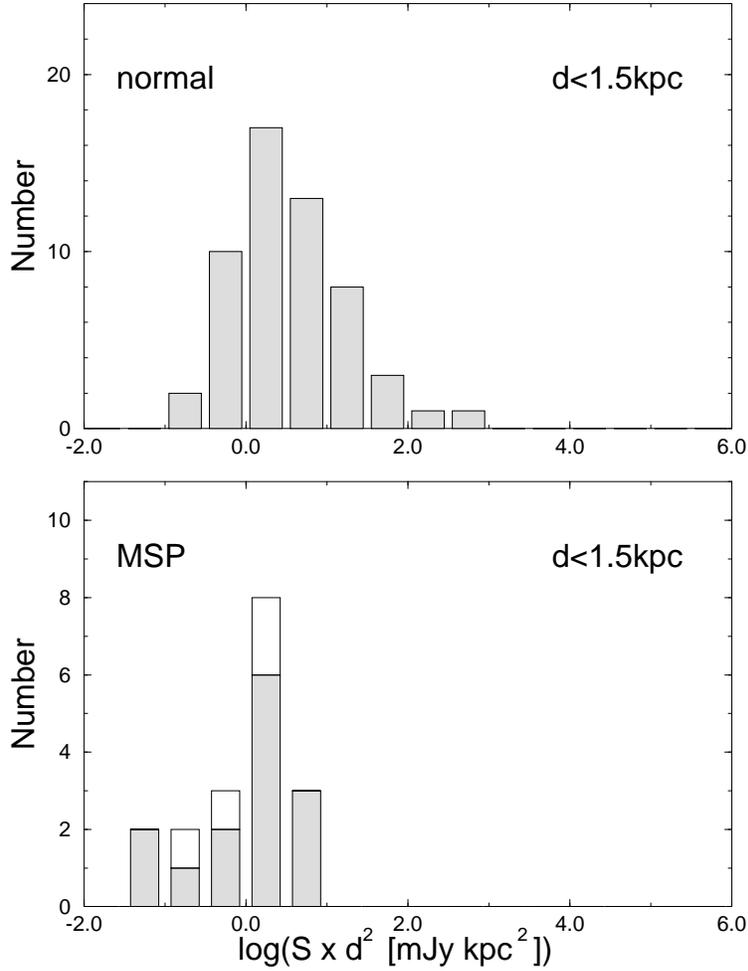}
\bigskip

\figcaption{\label{lum1400.lt.1.5} Luminosity distribution 
derived for sources within a distance of 1.5 kpc. Data 
for MSPs indicated by unshaded bars are derived from the literature
(see text for details). We note that there seems to be a particular
deficit in the number of high-luminosity MSPs relative to the normal
pulsars.}
\end{figure}

The study of the characteristics of MSP radio emission of which the
present work is the first of a series of papers,
was motivated by the question, in what respect do MSPs differ from normal
pulsars. The fact that both populations exhibit identical flux
density spectra points towards the same emission mechanism, which is
further supported by the results presented in paper II. At the same time,
the radio output of MSPs seems to be affected by their particular 
evolutionary history, i.e.,~the entire sample tends to be less
luminous and in fact less efficient radio emitters compared to normal
pulsars. It is particularly interesting that,
compared to the luminosity distribution of normal pulsars within a distance
of 1.5 kpc (Fig.~\ref{lum1400.lt.1.5}), we are apparently missing
some high luminosity MSPs, although they should be the easiest to detect.
Although no significant 
correlations between spectral parameters and intrinsic spin parameters
are observed, we note a weak tendency for old MSPs to exhibit flatter spectra
than MSPs with smaller intrinsic ages. 

Although remarkable exceptions to the rule are observed, the pulse 
profiles of MSPs are only slightly more complex than for the normal
pulsars. These profiles do not change
significantly with frequency (for a detailed discussion see
paper II), but we have indications that a measure of profile change is
actually related to the amount of mass transferred onto the 
neutron star during a spin-up episode, or to the
spin period. Moreover, the angular beam radii inferred from the 
observed pulse shapes imply that the emission beam of rapidly-rotating
pulsars is smaller than expected from our knowledge of normal pulsars.
If the neutron star equation-of-state allows the existence of sub-millisecond
pulsars, their detection will thus not be prevented by a beam size being too
large as it would have been implied from simply scaling the trend seen in
normal pulsars.

The present data  
suggest that, although MSPs and normal pulsars exhibit
the same emission physics, they show pronounced differences probably related
to the different evolutionary history. In paper
II we elaborate on these aspects  further and again raise the
question about the origin of these differences in light of the 
additional data. 

We strongly suggest that birth-rate calculations of MSP be
reconsidered given the data discussed here.

\acknowledgments 
We are grateful to the operators and engineers in Effelsberg for their support
during this project. It is also a pleasure to thank Norbert Wex for extremely
helpful discussions leading to the results presented in Sect.~4.3. We thank Alexis 
v.~Hoensbroech, Jarek Kijak 
and Christoph Lange for their help with the observations.
    FC gratefully acknowledges the support of the European Commission
    through a Marie Curie fellowship, under contract no. ERBFMBICT961700.
This work was in part supported by the European Commission
under the HCM Network Contract Nr.~ERB CHRX CT960633, 
i.e.,~the {\it European Pulsar Network}.

\newpage

\newpage

\begin{deluxetable}{lrcr@{$\; \pm \;$}lccl}
\small
\tablenum{1}
\tablewidth{0pt}
\tablecaption{\label{fluxtab} 
Flux densities and spectral indices for 34
millisecond pulsars. The 
flux density data presented for the first 23 sources
were derived from our measurements, while for 11 sources we collected
data 
from the literature. Besides the period of the pulsar (column 2),
we quote the corresponding frequency in column 3. The mean flux density
and its uncertainty is shown in column 4. Additionally, we present the
median of all measured flux values, which is to be compared with the mean
value. The spectral index is derived by using published lower frequency data
(for references see column 7) which is indicated in column 6.}
\tablehead{
\colhead{PSR} & \colhead{P} & \colhead{$\nu_{\rm eff}$} & 
   \multicolumn{2}{c}{$S_{\rm mean}$} & \colhead{$S_{\rm median}$} & 
 \colhead{spectral index}  & \colhead{Ref.} \\
 \colhead{}      & \colhead{(ms)} & \colhead{(GHz)} & \multicolumn{2}{c}{(mJy)} & 
   \colhead{(mJy)} & \colhead{$\alpha$} & \colhead{}  }
\startdata
J0613$-$0200 & 3.062 &  1.410  & 1.4\phn & 0.2 & 1.5\phn & $-1.9\pm0.4$ & 1 \nl
J0751+1807 &   3.479 &  1.435  & 3.2\phn & 0.7 & 2.9\phn & $-0.9\pm0.3$ & 2 \nl
J1012+5307 &   5.256 &  1.410  & 3\phd\phn\phn & 1   & 1.9\phn & $-1.8\pm0.4$ & 3 \nl
J1022+1001 &  16.453 &  1.443  & 3.0\phn & 0.4 & 2.5\phn & $-1.7\pm0.1$ & 4, 5 \nl 
J1024$-$0719 & 5.162 &  1.410  & 0.66 & 0.04 & 0.66 & $-1.5\pm0.2$ & 6 \nl
\tablevspace{5pt}
B1257+12   &   6.219 &  1.410  & 2\phd\phn\phn & 1  & 2\phd\phn\phn  & $-1.9\pm0.5$ & 7 \nl 
J1518+4904 &  40.935 &  1.410  & 4\phd\phn\phn & 2  & 2.5\phn & $-0.5\pm0.5$ & 26, 30 \nl
B1534+12   &  37.904 &  1.530  & 0.6\phn & 0.2 & 0.4\phn & $-3.0\pm0.4$ & 8 \nl
B1620$-$26 &  11.076 &  1.460  & 1.6\phn & 0.3 & 1.7\phn & $-2.5\pm0.2$ & 9, 10 \nl
J1640+2224 &   3.163 &  1.472  & 2\phd\phn\phn & 1 & 1\phd\phn\phn & $-0.3\pm0.5$ & 11 \nl
\tablevspace{5pt}
J1643$-$1224 & 4.621 &  1.410  & 4.8\phn & 0.6   & 4.5\phn  & $-2.4\pm0.3$ & 1 \nl
J1713+0747 &   4.570 &  1.510  & 8\phd\phn\phn & 2 & 5\phd\phn\phn & $-1.5\pm0.1$ & 12 \nl
J1730$-$2304 & 8.123 &  1.410  & 4\phd\phn\phn & 3   & 2\phd\phn\phn  & $-1.8\pm0.3$ & 1 \nl
J1744$-$1134 & 4.075 &  1.470  & 3\phd\phn\phn & 1 & 2\phd\phn\phn & $-1.6\pm0.3$ & 6 \nl
B1802$-$07 &  23.101 &  1.410  & 1.0\phn & 0.5 & 1.0\phn & $-1.0\pm0.5$ & 13 \nl
\tablevspace{5pt}
B1855+09   &   5.362 &  1.470  & 5\phd\phn\phn & 1 & 4\phd\phn\phn & $-1.3\pm0.2$ & 10, 14 \nl
B1913+16   &  59.030 &  1.560  & 0.9\phn & 0.2 & 0.9\phn & $-1.4\pm0.3$ & 15, 27 \nl
B1937+21   &  1.558  &  1.579  & 10\phd\phn\phn & 1 & 9\phd\phn\phn & $-2.3\pm0.2$ & 10 \nl
B1953+29   &  6.133  &  1.410  & 1.1\phn & 0.2 & 1.1\phn  & $-2.2\pm0.3$ & 16  \nl
J2051$-$0827 & 4.509 &  1.410  & 2.8\phn & 0.6 & 2.5\phn & $-1.8\pm0.3$ & 17 \nl
\tablevspace{5pt}
J2145$-$0750 & 16.052 & 1.510  & 8\phd\phn\phn & 2 & 7\phd\phn\phn & $-1.6\pm0.1$ & 18 \nl
J2229+2643 &   2.978 &  1.430  & 0.9\phn & 0.2 & 0.8\phn & $-2.2\pm0.5$ & 19 \nl
J2317+1439 &   3.445 &  1.496  & 4\phd\phn\phn & 1 & 2\phd\phn\phn & $-1.4\pm0.3$ & 20 \nl
\tablevspace{5pt}
\hline
\tablevspace{5pt}
J0218+4232 & 2.323 & 1.410 & 0.9\phn & \phn0.2 & $-$ & $-2.8\pm0.2$ & 21 \nl
J0437$-$4715 & 5.757 & 1.460 & 142\phd\phn\phn & 53 & $-$ & $-1.8\pm0.2$ & 22, 23 \nl
J0621+1002 & 28.853 & 0.800 & 1.9\phn & \phn0.3 & $-$ & $-2.5\pm1.4$ & 4 \nl
J0711$-$6830 & 5.491 & 1.400 & 1.6\phn & \phn0.3 & $-$ & $-1.4\pm0.2$ & 6 \nl
J1045$-$4509 & 7.474 & 1.520 & 3\phd\phn\phn & \phn1 & $-$ 
& $-1.5\pm0.3$ & 18 \nl
\tablevspace{5pt}
J1603$-$7202 & 14.841 & 1.400 & 3.0\phn & \phn0.5 & $-$ & $-1.3\pm0.3$ & 24 \nl
J1804$-$2717 & 9.343 & 1.400 & 0.4\phn & \phn0.2 & $-$ & $-3.2\pm0.5$ & 24 \nl
J1911$-$1114 & 3.625 & 1.400 & 0.5\phn & \phn0.2 & $-$ & $-3.3\pm0.4$ & 24 \nl
B1957+20 & 1.607 & 1.400 & 0.4\phn & \phn0.2 & $-$ & $-3.5\pm0.5$ & 25, 30 \nl
J2124$-$3358 & 4.931 & 1.400 & 1.6\phn & \phn0.4 & $-$ & $-1.1\pm0.4$ & 6 \nl
\tablevspace{5pt}
J2129$-$5721 & 3.726 & 1.400 & 1.4\phn & \phn0.2 & $-$ & $-1.0\pm0.4$ & 24 \nl
\enddata
\tablerefs{
1 - Lorimer et al.~(1995a), 2 - Lundgren et al.~(1995), 3 - Nichastro et 
al.~(1995), 4 - Camilo et al.~(1996b), 
5 - Kijak et al.~(1997a), 6 - Bailes et al.~(1997), 7 - Wolszczan \& 
Frail (1992), 8 - Wolszczan (1991), 
9 - Lyne et al.~(1988), 10 - Foster et al.~(1991), 11 - Foster et al.~(1995),
12 - Foster et al.~(1993), 
13 - D'Amico et al.~(1993), 14 - Segelstein et al.~(1986), 15 - Lorimer et 
al.~(1995b), 16 - Boriakoff et al.~(1983), 
17 - Stappers et al.~(1996), 18 - Bailes et al.~(1994), 19 - Camilo et 
al.~(1996a), 20 - Camilo et al.~(1993), 
21 - Navarro et al.~(1995), 22 - Johnston et al.~(1993), 23 - Manchester \& 
Johnston (1995), 
24 - Lorimer et al.~(1996), 25 - Fruchter et al.~(1988),
26 -  Nice et al.~(1996) 
28 - Taylor \& Weisberg (1982), 29 - Sayer et al.~(1997), 30 - Fruchter et al.~(1990)
}
\end{deluxetable}

\newpage

\begin{deluxetable}{lrrrrrrrrrr@{$\; \pm \;$}l}
\small
\tablenum{2}
\tablewidth{0pt}
\tablecaption{\label{widths} 
Pulse widths measured for millisecond pulsars. Besides 
the period of each pulsar (column 2) we quote the width at a 50\% and
10\% level  and their corresponding uncertainty in degrees (columns 3 -- 6),
and in units of time (columns 7 -- 10). We used the 10\% width to derive a duty
cycle quoted in column 11.}
\tablehead{
\colhead{PSR} & \colhead{P} & \colhead{$W_{\rm 50}$} & \colhead{$W_{\rm 10}$} & 
\colhead{$W_{\rm eq}$} & \colhead{$\Delta W$} & \colhead{$W_{\rm 50}$}        &
\colhead{$W_{\rm 10}$} & \colhead{$W_{\rm eq}$} & \colhead{$\Delta W$}        & 
\multicolumn{2}{c}{10\% Duty Cycle} \\
\colhead{}   & \colhead{(ms)} & \colhead{(deg)} & \colhead{(deg)}     & 
\colhead{(deg)} & \colhead{(deg)} & \colhead{(ms)} & \colhead{(ms)}   & 
\colhead{(ms)} & \colhead{(ms)} & \multicolumn{2}{c}{(\%)}              }
\startdata
J0613$-$0200 &   3.062 & 101\phd\phn   & 120\phd\phn &  58\phd\phn &   9\phd\phn &   0.86\phn &   1.02\phn &   0.49\phn &  0.08\phn &  33\phd\phn\phn   &    3\phd\phn\phn \nl
J0621+1002 &  28.854 & 114.2 & 124.0 &  20.9 &   0.9 &   9.15\phn &   9.94\phn &   1.68\phn &  0.07\phn &  34.4\phn &  0.3\phn \nl
J0751+1807 &   3.479 &  73\phd\phn   & 101\phd\phn &  42\phd\phn &   6\phd\phn &   0.70\phn &   0.98\phn &   0.40\phn &  0.06\phn &  28\phd\phn\phn   &    2\phd\phn\phn \nl
J1012+5307 &   5.256 &  47\phd\phn   &  69\phd\phn &  41\phd\phn &   1\phd\phn &   0.69\phn &   1.01\phn &   0.59\phn &  0.02\phn &  19.2\phn &  0.3\phn \nl
J1022+1001 &  16.453 &  23.3 &  43.2 &  22.8 & 0.4 &   1.07\phn &   1.97\phn &   1.04\phn &  0.02\phn &  12.0\phn &  0.1\phn \nl
\tablevspace{5pt}
J1024$-$0719 &   5.612 &  41.1 & 111.9 & 46.8 &   0.9 &   0.59\phn &   1.60\phn &   0.67\phn &  0.01\phn &  31.1\phn &  0.3\phn \nl
B1257+12   &   6.219 &  38\phd\phn   &  77\phd\phn &  44\phd\phn &   1\phd\phn &   0.66\phn &   1.34\phn &   0.75\phn &  0.02\phn &  21.5\phn &  0.3\phn \nl
J1518+4904 &  40.935 &  12.5 &  19.3 &  11.1 &   0.6 &   1.42\phn &   2.19\phn &   1.26\phn &  0.07\phn &   5.4\phn &  0.2\phn \nl
B1534+12   &  37.904 &   2.9 &  24.1 &   \phn3.7 &   0.6 &   0.31\phn &   2.54\phn &   0.39\phn &  0.06\phn &   6.7\phn &  0.2\phn \nl
B1620$-$26 &  11.076 &  67\phd\phn &  88\phd\phn &  22\phd\phn &   4\phd\phn &   2.1\phn\phn  &   2.7\phn\phn  &   0.7\phn\phn  &  0.1\phn\phn  &  25\phd\phn\phn   &  1\phd\phn\phn   \nl
\tablevspace{5pt}
J1640+2224  &   3.163 &  25\phd\phn &  56\phd\phn &  27\phd\phn &   4\phd\phn &   0.22\phn &   0.50\phn &   0.24\phn &  0.04\phn &  16\phd\phn\phn   &  1\phd\phn\phn   \nl
J1643$-$1224 & 4.621 &  32\phd\phn &  78\phd\phn &  41\phd\phn &   2\phd\phn &   0.41\phn &   1.00\phn &   0.52\phn &  0.03\phn &  21.6\phn &  0.6\phn \nl
J1713+0747 &   4.570 &  11\phd\phn & 100\phd\phn &  15\phd\phn &   3\phd\phn &   0.14\phn &   1.27\phn &   0.19\phn &  0.03\phn &  27.8\phn &  0.7\phn \nl
J1730$-$2304 & 8.123 &  47.4 &  77.0 &  39.1 &   0.9 &   1.07\phn &   1.73\phn &   0.88\phn &  0.02\phn &  21.4\phn &  0.3\phn \nl
J1744$-$1134 & 4.075 &  13.8 &  26.8 &  15.2 &   0.6 &  0.156 &  0.303 &  0.172 &  0.007&  7.4  &  0.2 \nl
\tablevspace{5pt}
B1744$-$24A & 11.564 &  26\phd\phn &  47\phd\phn &  28\phd\phn &  15\phd\phn &   0.8\phn\phn  &   1.5\phn\phn  &    0.9\phn\phn &  0.5\phn\phn  &  13\phd\phn\phn   &   4\phd\phn\phn  \nl
B1802$-$07 &  23.101 &  57\phd\phn &  82\phd\phn &  30\phd\phn &   6\phd\phn &   3.7\phn\phn  &   5.3\phn\phn  &   1.9\phn\phn  &  0.4\phn\phn &  23\phd\phn\phn &  2\phd\phn\phn \nl
B1855+09   &   5.362 &  36\phd\phn &  54\phd\phn &  34\phd\phn &   2\phd\phn &   0.54\phn &   0.80\phn &   0.51\phn &  0.03\phn &  14.9\phn &  0.5\phn \nl
B1913+16   &  59.030 &  45\phd\phn &  52\phd\phn &  21\phd\phn &   2\phd\phn &   7.4\phn\phn  &   8.5\phn\phn  &   3.5\phn\phn  &  0.3\phn\phn  &  14.3\phn &  0.6\phn \nl
B1937+21\tablenotemark{a} &
      1.558 &  14.6 &  20.0 &  10.5 &   0.2 &  0.063 &  0.087 &  0.045 &  0.001 &   5.55 &  0.06 \nl
\tablevspace{5pt}
B1953+29   &   6.133 & 107\phd\phn & 131\phd\phn &  60\phd\phn &  12\phd\phn &   1.8\phn\phn  &   2.2\phn\phn  &   1.0\phn\phn  &  0.2\phn\phn  &  36\phd\phn\phn   &   3\phd\phn\phn  \nl
J2019+2425 &   3.935 &  67\phd\phn &  96\phd\phn &  62\phd\phn &   3\phd\phn &   0.73\phn &   1.05\phn &   0.68\phn &  0.03\phn &  26.7\phn & 0.8\phn \nl
J2051$-$0827 & 4.509 &  27\phd\phn &  107\phd\phn & 32\phd\phn & 3\phd\phn &   0.34\phn &   1.33\phn &   0.40\phn &  0.04\phn &  29.6\phn &  0.9\phn \nl
J2145$-$0750 & 16.052 & 88.6 &  98.4 &  23.6 &   0.4 &   3.95\phn &   4.39\phn &   1.05\phn &  0.02\phn &  27.3\phn &  0.1\phn \nl
J2229+2643 &   2.978 &  70\phd\phn &  88\phd\phn &  53\phd\phn &   6\phd\phn &   0.58\phn &   0.73\phn &   0.44\phn &  0.05\phn &  24\phd\phn\phn   &  2\phd\phn\phn   \nl
\tablevspace{5pt}
J2317+1439 &   3.445 &  48\phd\phn &  74\phd\phn&  48\phd\phn &   5\phd\phn &   0.46\phn &   0.71\phn &   0.45\phn &  0.04\phn &  21\phd\phn\phn   &  1\phd\phn\phn   \nl
J2322+2057 &   4.808 &  23\phd\phn &  51\phd\phn &  27\phd\phn &   2\phd\phn &   0.30\phn &   0.68\phn &   0.36\phn &  0.03\phn &  14.2\phn &  0.6\phn \nl
\enddata
\tablenotetext{a}{width measured from EBPP data (see text for details)}
\end{deluxetable}

\newpage

\begin{deluxetable}{lrr@{$\; \pm \; $}lc}
\small
\tablenum{3}
\tablewidth{0pt}
\tablecaption{\label{rhotab} Opening angles for MSPs derived from  
profile widths measured at 10\% intensity level. Calculations are based
on fits of the rotating-vector-model to polarisation data to derive
the viewing geometry (paper IV). In cases where this is
not possible, we quote values expected from statistical arguments, 
$\langle \rho_{10} \rangle$
(see text for details).}
\tablehead{
\colhead{PSR} & \colhead{P} & \multicolumn{2}{c}{$\rho_{10}$} 
& \colhead{$\langle \rho_{10} \rangle$} \\
\colhead{}   & \colhead{(ms)} & \multicolumn{2}{c}{(deg)} & \colhead{(deg)} }
\startdata
J0613$-$0200 &   3.062 & \multicolumn{2}{c}{$-$} &  61.5 \nl
J0621+1002 &  28.854 &  \multicolumn{2}{c}{$-$} &  62.4 \nl
J0751+1807 &   3.479 &  \multicolumn{2}{c}{$-$} & 57.2 \nl
J1012+5307 &   5.256 &  35\phd\phn & 1\phd\phn & $-$  \nl
J1022+1001 &  16.453 &  25.0 &  0.6 & $-$ \nl
\tablevspace{5pt}
J1024$-$0719 &   5.612 & 25.8 & 0.1 & $-$ \nl
B1257+12   &   6 .219 &  40.0 & 0.9 & $-$ \nl
J1518+4904 &  40.935 &  10.5 &  0.4 & $-$ \nl
B1534+12\tablenotemark{a} &  37.904 &   17\phd\phn &  5 & $-$ \nl
B1620$-$26 &  11.076 &  46\phd\phn & 3\phd\phn & $-$ \nl
\tablevspace{5pt}
J1640+2224  &   3.163 &  6.9 & 0.1 & $-$ \nl
J1643$-$1224 & 4.621 &  23\phd\phn &  1\phd\phn &  $-$ \nl
J1713+0747 &   4.570 &  54\phd\phn & 3\phd\phn &  $-$ \nl
J1730$-$2304 & 8.123 &  \multicolumn{2}{c}{$-$} &  52.0 \nl
J1744$-$1134 & 4.075 &  18.1 & 0.4 & $-$ \nl
\tablevspace{5pt}
B1744$-$24A & 11.564 &  \multicolumn{2}{c}{$-$} & 46.1 \nl
B1802$-$07 &  23.101 &  \multicolumn{2}{c}{$-$} & 53.1 \nl
B1855+09   &   5.362 &  27\phd\phn &  2\phd\phn & $-$ \nl
B1913+16\tablenotemark{b} &  59.030 &  21\phd\phn &  4\phd\phn & $-$ \nl
B1937+21  & 1.558 &   \multicolumn{2}{c}{$-$} & 35.8 \nl
\tablevspace{5pt}
B1953+29   &   6.133 & \multicolumn{2}{c}{$-$} & 63.9 \nl
J2019+2425 &   3.935 &  \multicolumn{2}{c}{$-$} & 56.1 \nl
J2051$-$0827 & 4.509 &  \multicolumn{2}{c}{$-$} & 58.6 \nl
J2145$-$0750 & 16.052 & \multicolumn{2}{c}{$-$} & 56.7 \nl
J2229+2643 &   2.978 &  25\phd\phn &  3\phd\phn & $-$ \nl
\tablevspace{5pt}
J2317+1439 &   3.445 &  10\phd\phn &  1\phd\phn & $-$ \nl
J2322+2057 &   4.808 &  \multicolumn{2}{c}{$-$} & 43.3 \nl
\enddata
\tablenotetext{a}{using viewing geometry derived by Arzoumanian et al.~(1996)}
\tablenotetext{b}{using viewing geometry constrained by Cordes et al.~(1990)}
\end{deluxetable}

\end{document}